\documentclass[review]{elsarticle}

\usepackage{graphicx}
\usepackage{multicol}
\usepackage{amsmath}
\usepackage{amsfonts, amssymb}
\usepackage{multirow}
\usepackage{rotating}
\usepackage{marvosym}
\usepackage{caption}
\usepackage{subcaption}
\usepackage{array}

\usepackage[hidelinks]{hyperref}

\usepackage[colorinlistoftodos, textwidth=40mm, textsize=footnotesize]{todonotes}

\newcommand{\AF}[1]{#1}

\newtheorem{remark}{Remark}

\newcommand{\bs}[1]{\boldsymbol{#1}}
\newcommand{\T}{^\top}
\renewcommand{\th}{\text{-th}}
\newcommand{\on}{{\text{ \Lightning}}}
\newcommand{\dt}{{\Delta}}
\newcommand{\du}{{\Delta_u}}
\newcommand{\ch}{{ch}}
\newcommand{\mt}{{mt}}

\newcommand{\B}{{\mathcal B}}
\newcommand{\C}{{\mathcal C}}
\newcommand{\St}{{\mathcal S}}
\newcommand{\CHP}{{\mathcal M}}

\newcommand{\BigM}{{\mathbf M}}

\begin{document}

\begin{frontmatter}

\title{A compositional modeling framework for the optimal energy management of a district network}

\author[DEIB]{Daniele~Ioli}
\author[DEIB]{Alessandro~Falsone}
\author[MDH]{Alessandro~Vittorio~Papadopoulos\corref{CA}}
\author[DEIB]{Maria~Prandini}

\cortext[CA]{Corresponding author, Tel. +46 (0)21-1073 23}
\address[DEIB]{Politecnico di Milano, Piazza Leonardo da Vinci, 32, 20133 Milano, Italy}
\address[MDH]{M{\"a}lardalen University, H{\"o}gskoleplan 1, 72123, V{\"a}ster{\aa}s, Sweden}
\fntext[myfootnote]{This work is partly supported by the European Commission under the UnCoVerCPS project, grant number 643921, and was performed when the third author was a post-doctoral researcher at Politecnico di Milano. }

\begin{abstract}
This paper proposes a compositional modeling framework for the optimal energy management of a district network. The focus is on cooling of buildings, which can possibly share resources to the purpose of reducing maintenance costs and using devices at their maximal efficiency. Components of the network are described in terms of energy fluxes and combined via energy balance equations. Disturbances are accounted for as well through their contribution in terms of energy. Different district configurations can be built, and the dimension and complexity of the resulting model will depend on the number and type of components and on the adopted disturbance description. Control inputs are available to efficiently operate and coordinate the district components, thus enabling energy management strategies to minimize the electrical energy costs or track some consumption profile agreed with the main grid operator. 
 \end{abstract}


\end{frontmatter}


\section{Introduction}

Building energy management, and temperature regulation in particular, has recently attracted the attention of various researchers (see, e.g.,~\cite{PuttaZKHB_HPBC12,PuttaKCHB_IBO13, PuttaKCHB_HPBC14b,Kontes2014846,Zavala2014714,wetter2014jbps,ISGT13,ECC2013,Scherer2014740,Morosan20101445,MaBorrelliHenceyCoffeyBengeaHaves10,Henze_etal_2005,MaKelmanDalyBorrelli12}).
Indeed, energy consumption in buildings represents approximately 40\% of the worldwide energy demand, and more than half of this amount is spent for Heating, Ventilation and Air Conditioning (HVAC) systems~\cite{PerezLombard2008394,book2012,MTNS2016}.
Energy management can be performed at the level of a single building, e.g., using the storage to shift in time the thermal energy request to time slots where the electricity costs are lower. As buildings started sharing equipments at the benefit of shared operating costs, increased flexibility, and overall performance improvement, energy management needs to be performed at the district network level, which calls for appropriate modeling and control strategies. Constructing models of interconnected systems is generally demanding, and here we propose a modular framework that simplifies this task and is also suitable for the application of different control design approaches.

The proposed modeling approach is oriented to energy management and compositional in that components are described in terms of thermal/electrical energy fluxes  and interact by exchanging energy, which makes easy to compose a district network configuration via energy balance equations.
Our modeling framework is built with a control-oriented perspective. It includes disturbances like, e.g., solar radiation, outside temperature, occupancy, and wind power production, as well as control inputs like, e.g., buildings temperature set-points, charge/discharge of storages, activation/deactivation of devices, that can be appropriately set so as to optimize performance at the district level.

Complexity and size of the model associated with a district configuration depend on number of components and type of description adopted per component. The model can be either deterministic or stochastic depending on the disturbance characterization as a deterministic or stochastic process, respectively. It can range from a low dimensional deterministic system  with continuous input and state that is convex in the control input, to a large dimensional  Stochastic Hybrid System (SHS)~\cite{LygerosPrandini10} with discrete and continuous input and state.

Given a certain configuration, one can then formulate energy management problems like the minimization of the cost of the electrical energy requested to the main grid or the tracking of some given electrical energy exchange profile that was agreed with the main grid operator according to a demand-response strategy. The district network in the latter case can be viewed as a user that actively participates to the electrical energy demand/generation balance of the overall grid, and, hence, to its stabilization.

\AF{Further contributions in the literature adopt a similar perspective. In \cite{lauster2015}, the focus is on simulation so that the model dependence on the control input is not a concern. In \cite{Korkas2015}, the aim is the design of an energy management strategy via a simulation-based approach. The modeling effort is limited in this case, and the idea is to take an accurate model in the literature and run simulations to the purpose of policy design, with no concern of making explicit the dependence on the input and formally proving optimality.
The approach in \cite{Nguyen2014} is the closest to our approach, in that it addresses energy management problems for a  microgrid that is built based on models of single components, combined via energy balance equations. Models are however simplified, in particular that of the building. Also, occupancy is not accounted for explicitly. A specific strategy for energy management is considered, whereas our framework is more comprehensive since it allows for the design of different strategies (certainty equivalence based, robust, stochastic) for the minimization of suitably defined (nominal, min-max, average) cost in presence of (nominal,  robust, probabilistic) constraints on comfort and actuation. Depending on the  network communication and computation capabilities, and on privacy issues like in the case of buildings not willing to disclose their consumption profile, a centralized, decentralized, or distributed optimization scheme can be conceived and implemented. Overall, our work is more general and it actually subsumes the approach in \cite{Nguyen2014}.}

It is worth noticing that other modeling frameworks have been developed in the literature~\cite{wetter2014jbps,bonvini2011mcmds,Wetter2016290}. However, the obtained models are typically more complex since they are based on partial differential equations, and require numerical optimization tools for solving the resulting nonlinear optimization problems~\cite{Wetter2016290,Zavala20091725,Gouda20021255,Shaikh2014409}.

This paper is based on our earlier work in~\cite{ioli_adchem2015,ioli2016icca}, which is extended in several directions. We provide a more detailed description of the district components, including a validation with respect to other commercial simulation tools of the building thermal model according to a norm defined by the American Society for Heating Refrigerating and Air-conditioning Engineers (ASHRAE). We show how to compose a network configuration and formulate an energy management problem as an optimization program. In particular, we show a simulation study of the results achieved in the case where nominal disturbances are present and computations are performed by a central unit. Finally, we suggest a multirate approach as a viable solution for allowing real-time computation of the control input, while retaining model accuracy.

The reminder of the paper is structured as follows. Section~\ref{sec:components} presents the models of the district network components, and Section~\ref{sec:configurations} shows how they can be connected to set up a network configuration, while defining objective and constraints of the optimal energy management problem. Section~\ref{sec:simResults} describes some configurations, providing examples of the kind of results one can achieve through the presented framework. Section~\ref{sec:multirate} shows how to deal with computational complexity, discussing a multirate approach, while Section~\ref{sec:conclusion} concludes the paper. \ref{sec:validation} describes the procedure adopted for validating the model of the building.

 \section{District network components}
\label{sec:components}

We consider a district network connected to the main grid that will provide the electrical energy needed to compensate for possible imbalance between demand and generation within the district. We model the evolution of the network over a finite time horizon $[t_i,t_f]$, which is divided into $M$ time slots of duration $\dt$. The contribution in terms of energy requested/provided by the different components per time slot along the discretized control horizon is provided.
Components can consume (e.g., buildings), produce (e.g., renewable power generators), store (e.g. thermal storages and batteries), or convert energy (e.g., the chiller plants), and are combined via energy balance equations so as to build the overall model of the district. Each component may be affected by some inputs which can be either disturbances or control inputs. In the case when control inputs are available, a suitable strategy can be conceived to set them so as to efficiently manage the system along the time horizon $[t_i,t_f]$.

In the rest of this section, we provide a model for the following components: \emph{building},  \emph{chiller}, \emph{storage}, \emph{combined heat and power unit}, and \emph{wind turbine}. Models are either derived from first principles or taken from the literature.  In the latter case, appropriate references are provided. Tables~\ref{Tab:summaryBuilding}--\ref{Tab:summaryCHP} summarize the main characteristics of the first 4 components. The last component provides an input to the network in terms of wind energy. \AF{Similarly to the wind energy contribution, one could consider the solar energy contribution provided by photovoltaic panel installations. Models partly derived from first principle and partly taken from the literature could be used to this purpose. This is not treated here, but the interested reader can refer to, e.g.,  \cite{ARCH2017}. Further components could also be added to the district network. The key idea when introducing our compositional framework is that if a component can be modeled in terms of energy, possibly depending on some control input and/or disturbance signal, then, it can be easily included in the network. When the dependence of the energy on the control input is convex, piecewise linear, or linear with additional binary variables, the problem of designing an energy management strategy can be reduced to a mixed integer linear or a convex optimization program for which efficient solvers exist. }

\subsection{Building}\label{sec:model_building}
We consider a building as composed of $n_z$ zones, where each zone is characterized by its own (average) temperature $T_{z,j}$, $j=1,\dots,n_z$.
The zones temperatures can be collected in a vector $\bs{T_z} = [T_{z,1} \cdots T_{z,n_z}]\T$ and we next determine the amount of cooling energy $E_c$ needed for making them track a given profile. We say that the building is controllable if a control layer is present to this purpose. Suitable constraints will be imposed on the assigned profile to make the resulting tracking problem feasible while guaranteeing comfort conditions at the same time.

The cooling energy $E_{c,j}$ requested by zone $j$ can be derived based on the thermal energy balance within the zone, accounting for both thermal effects related to its structure and thermal phenomena related to occupancy, equipment, lights, etc, and solar radiation through windows.
More precisely, we have
\begin{align}
	E_{c,j} = E_{w,j} + E_{z,j}+ E_{p,j} + E_{\mathrm{int},j},
\label{eq:cooling_energy_zone_j}
\end{align}
where $E_{w,j}$  is the amount of energy exchanged between  zone $j$ and its adjacent walls, $E_{z,j}$ is the contribution of the thermal inertia of zone $j$, and $E_{p,j}$ and $E_{\mathrm{int},j}$ is the heat produced by people and other heat sources within zone $j$, respectively.

The thermal model of the building is derived from first principles, following \cite{KimAndBrown,Kim2}.

\subsubsection{Walls contribution}
For modeling the walls contribution we use a one-dimensional finite volumes model. Each wall is divided into vertical layers (`slices') that may differ in width and material composition. The area of each slice coincides with the wall area and each slice is assumed to have uniform density and temperature. The one-dimensional discretization is sensible since the heat flow is perpendicular to the crossed surface. Each internal slice exchanges heat only with nearby slices through conduction, whilst boundary slices are exposed towards either a zone or the outside of the building and exchange heat also via convection and thermal radiation. External surfaces are assumed to be gray and opaque, with equal absorbance and emissivity and with zero transmittance. Absorbance and emissivity are wavelength-dependent quantities, and here we shall consider two different values for shortwave and longwave radiation.

The heat transfer balance equation for the $i\th$ slice \AF{of the $w\th$ wall is given by:
\begin{align}
	\dot{T}_{w,i} =& \frac{1}{C_{w,i}} \Big[ (k_{w,i}^{i-1} + h_{w,i}^{i-1})T_{w,i-1} + (k_{w,i}^{i+1} + h_{w,i}^{i+1})T_{w,i+1} \nonumber \\
		&-(k_{w,i}^{i-1} + h_{w,i}^{i-1} + k_{w,i}^{i+1} + h_{w,i}^{i+1})T_{w,i} + Q_{g,w,i} + R_{w,i} \Big], \label{eq:slice_balance}
\end{align}
where $T_{w,i}$ denotes the temperature of the wall slice, $C_{w,i}$ being its thermal capacity per unit area, and $k_{w,i}^j$ and $h_{w,i}^j$, with $j = i\pm 1$, representing respectively the conductive and convective heat transfer coefficients between the $i^\mathrm{th}$ and the $j^\mathrm{th}$ slice of the same wall $w$. $Q_{g,w,i}$ is the thermal power generation inside slice $i$ and $R_{w,i}$ represents radiative heat exchanges and is defined as
\begin{equation*}
	R_{w,i} =\hspace{-.2em}  \begin{cases}
		0,		&\hspace{-1em} 1 < i < m \\
		\alpha_{w}^S Q^S + \alpha_{w}^L Q^L - \varepsilon_{w,i} Q_r(T_{w,i}),	&
\hspace{-.5em} \text{slice $i$ facing outside} \\
		\displaystyle \hspace{-.2em} \sum_{\substack{w'=1,\dots,n_w\\ j\in\{1,M\}}} \hspace{-1.3em} F_{(w,i)\to(w',j)} \left( \varepsilon_{w',j} Q_r(T_{w',j}) - \varepsilon_{w,i} Q_r(T_{w,i}) \right)	 &\hspace{-0.5em} \text{slice $i$ facing inside}
	\end{cases}
\end{equation*}
where $Q^S$ and $Q^L$ denote the incoming shortwave and longwave radiation power per unit area, respectively, and $\alpha_{w}^S$ and $\alpha_{w}^L$ are the corresponding absorbance rates for wall $w$. $Q_r(T_{w,i})$ is the emitted radiation as a function of the slice temperature, $\varepsilon_{w,i} < 1$ being the emissivity and $F_{(w,i)\to(w',j)}$ the view factor that takes into account the fraction of radiation leaving slice $i$ of wall $w$ and reaching slice $j$ of wall $w'$. Finally, $n_w$ denotes the total number of walls.

Equation \eqref{eq:slice_balance} holds for every slice in every wall $w$.
If the wall is composed of $m$ slices, we have $m$ equations like \eqref{eq:slice_balance} with $i=1,2,\dots,m$.
When the superscript in the right-hand side of equation \eqref{eq:slice_balance} takes value $0$ or $m+1$, reference is made to either a zone of the building (internal surface of the wall) or the outside of the building (external surface of the wall). Note that $k_{w,1}^0 = k_{w,m}^{m+1} = 0$ as there is no thermal conduction on walls boundary surfaces, $h_{w,i}^{i-1} = 0$ for $i>1$, $h_{w,i}^{i+1} = 0$ for $i<m$, and $\varepsilon_{w,i} = 0$ for $1<i<m$, since there is no thermal convection nor radiation between slices. As for the slice in contact with the ground, we assume that the energy exchange occurs via thermal conduction only (no convection nor radiation is considered), where the ground is considered as a thermal reservoir, and as such its temperature is constant.
Since we assume that each wall is a gray body, the power $Q_r(T_{w,i})$ radiated from each slice is governed by $Q_r(T_{w,i}) = \sigma T_{w,i}^4$, where $\sigma$ is the Stefan-Boltzmann constant. This expression is approximately linear around the slice mean operating temperature $\overline{T}_{w,i}$ so that it can be replaced by
\begin{align}
	Q_r(T_{w,i}) = 4 \sigma \overline{T}_{w,i}^3 T_{w,i} - 3 \sigma \overline{T}_{w,i}^4.
	\label{eq:stefan_boltzmann_lin}
\end{align}
Then, the evolution of the temperatures $\bs{T_w} = [T_{w,1} \cdots T_{w,m}]\T$ of the $m$ slices composing wall $w$ can be described in matrix form by
\begin{align}
	\bs{\dot{T}_w} = \bs{A_w}\bs{T_w} + \bs{B_w}\bs{T_z} + \bs{W_w}\bs{d},
	\label{eq:wall_dynamics}
\end{align}
where we recall that $\bs{T_z}$ is the vector containing the temperatures of the $n_z$ zones.
Vector $\bs{d} = [T_{out}\; T_{gnd} \; Q^S \; Q^L \; 1]\T$ is the disturbance input and collects the outdoor temperature $T_{out}$, the ground temperature $T_{gnd}$, and the incoming shortwave $Q^S$ and longwave $Q^L$ radiations. The constant $1$ in $\bs{d}$ is introduced to account for the constant term in \eqref{eq:stefan_boltzmann_lin}. Finally, $\bs{A_w}$, $\bs{B_w}$ and $\bs{W_w}$ are suitably defined matrices that are easily derived based on the scalar equation \eqref{eq:slice_balance}, whose coefficients depend on the wall characteristics.
}

Equation \eqref{eq:wall_dynamics} refers to a single wall. If there are $n_w$ walls in the building, then, we can collect all walls temperatures in vector $\bs{T} = [\bs{T_1\T} \cdots \bs{T_{n_w}\T}]\T$, and write the following equation for the evolution in time of $\bs{T}$:
\begin{equation}
	\bs{\dot{T}} = \bs{A}\bs{T} + \bs{B}\bs{T_z} + \bs{W}\bs{d},
	\label{eq:building_dynamics}
\end{equation}
where $\bs{A}$ is a block-diagonal matrix with $\bs{A_w}$ as $w\th$ block, $\bs{B} = \begin{bmatrix}\bs{B_1\T} & \cdots & \bs{B_{n_w}\T}\end{bmatrix}\T$ and
$\bs{W} = \begin{bmatrix}\bs{W_1\T} \cdots \bs{W_{n_w}\T}\end{bmatrix}\T$.

If we consider zone $j$ and one of its adjacent wall $w$, then the thermal power transferred from wall $w$ to zone $j$ is given by
\begin{align}\label{eq:w-z}
	Q_{w \rightarrow j} = S_w h_{w,b}^{b'} (T_{w,b} - T_{z,j}),
\end{align}
where $S_w$ is the wall surface and the pair $(b,b')$ can be either  $(1,0)$ or $(m,m+1)$ according to the notation introduced for \eqref{eq:slice_balance}.  The total amount of thermal power transferred from the building walls to zone $j$ can be expressed as $Q_{b,j} = \sum_{w\in \mathcal{W}_j} Q_{w\rightarrow j}$, where $\mathcal{W}_j$ is the set of walls $w$ adjacent to zone $j$. Defining $\bs{Q} = [Q_{b,1} \cdots Q_{b,n_z}]\T$, we obtain
\begin{equation}
	\bs{Q} = \bs{C}\bs{T} + \bs{D}\bs{T_z},
	\label{eq:building_output}
\end{equation}
where $\bs{C}$ and $\bs{D}$ are suitably defined matrices derived based on equation \eqref{eq:w-z}. From \eqref{eq:building_dynamics} and \eqref{eq:building_output}, we finally get
\begin{equation}
\begin{cases}
\bs{\dot{T}} = \bs{A}\bs{T} + \bs{B}\bs{T_z} + \bs{W}\bs{d} \\
\bs{Q} = \bs{C}\bs{T} + \bs{D}\bs{T_z}
\end{cases}
\label{eq:building_state_space}
\end{equation}

\begin{remark}
The obtained model, though linear, can be quite large. However,  its order can be greatly reduced by applying the model reduction algorithm
based on Hankel Single Value Decomposition (HSVD), as suggested in \cite{KimAndBrown}.
\hfill \ \qed
\end{remark}

The zone temperature profile to track $\bs{T_z}$ is taken as a linear function of time within each time slot of length $\dt$, defined by the values $u(k) = \bs{T}_{\bs{z}}(k\dt)$ at the time steps $k=0,1,\dots,M$.
By approximating the input $\bs{d}$ as a piecewise linear function of time as well, with values $\omega(k) = \bs{d}(k\dt)$ at $k=0,1,\dots,M$, an exact discrete time version of the linear model \eqref{eq:building_state_space} can be derived (see~\ref{sec:discretization_walls}).
The evolution of $y(k) = \bs{Q}(k\dt)$ over the finite time horizon can then be computed as
\begin{equation}
\bs{y} = [y\T(0) \cdots y\T(M)]\T= F\bs{T}(0) + G\bs{u} + H\bs{\omega}
	\label{eq:discrete_output}
\end{equation}
where we set $\bs{u} = [u\T(0) \cdots u\T(M)]\T$ and $\bs{\omega} = [\omega\T(0) \allowbreak \cdots \allowbreak \omega\T(M)]\T$,
and $F$, $G$ and $H$ are suitably defined matrices.

The thermal energy $E_w(k) = [E_{w,1}(k) \cdots E_{w,n_z}(k)]\T$ transferred from the walls to all zones can be computed by integrating $\bs{Q}(t)$ on each time
slot, which leads to the following approximate expression:
\begin{equation}\label{eq:wall_energy_tot}
	E_w(k) = \frac{\dt}{2} ( y(k-1) + y(k) ), k=1,\dotsc,M.
\end{equation}
Finally, from \eqref{eq:discrete_output} and \eqref{eq:wall_energy_tot} we can derive the enlarged energy vector $\bs{E_w} = [E_w\T(1) \cdots E_w\T(M)]\T$:
\begin{equation}
	\bs{E_w} = \tilde{F}\bs{T}(0) + \tilde{G}\bs{u} + \tilde{H}\bs{\omega},
	\label{eq:wall_energy_tot_matrix}
\end{equation}
where $\tilde{F}$, $\tilde{G}$, and $\tilde{H}$ are obtained from matrices ${F}$, ${G}$, and ${H}$ in \eqref{eq:discrete_output} via \eqref{eq:wall_energy_tot}.

\subsubsection{Zones energy contribution}

In order to decrease the temperature of zone $j$ in the time frame from $(k-1)\dt$ to $k\dt$, we need to draw energy from the zone itself. This energy contribution can be expressed as
\begin{equation}
	E_{z,j}(k) = - C_{z,j} (T_{z,j}(k\dt) - T_{z,j}((k-1)\dt)),
\label{eq:zones_energy}
\end{equation}
where $C_{z,j}$ is the heat capacity of the $j\th$ zone.
If we account for all $n_z$ zones, and all $M$ time frames within the finite horizon $[t_i,t_f]$, equation \eqref{eq:zones_energy} can be written in the following matrix form
\begin{equation}
\bs{E_z} = Z\bs{u},
\label{eq:zone_energy_equation}
\end{equation}
where we set $\bs{E_z} = [E_z\T(1) \cdots E_z\T(M)]\T$ with $E_z(k) = [E_{z,1}(k) \allowbreak \cdots \allowbreak E_{z,n_z}(k)]\T$,
and $Z$ is a suitably defined matrix.

\subsubsection{People energy contribution}
Occupancy implies heat production, which in crowded places can be actually significant~\cite{Zavala2014714}.
According to an empirical model in \cite{CIBSEGuideA}, the heat rate $Q_{p,j}$ produced by the $n_{p,j}$ occupants of a zone $j$ at temperature $T_{z,j}$ is given by
\begin{equation}
	Q_{p,j} = n_{p,j} ( p_2 T_{z,j}^2 + p_1 T_{z,j} + p_0 ),
	\label{eq:people_heat_rate}
\end{equation}
where $p_2 = -0.22$\,W/K$^2$, $p_1 = 125.12$\,W/K and $p_0 = -1.7685 \cdot 10^4$\,W. Expression \eqref{eq:people_heat_rate} is almost linear in a sensible operating temperature range and can thus be accurately approximated by linearization around some comfort temperature $\overline{T}_{z,j}$:
\begin{align}
Q_{p,j} &= n_{p,j} \left( (2 p_2 \overline{T}_{z,j} + p_1)(T_{z,j}-\overline{T}_{z,j}) + p_2 \overline{T}_{z,j}^2 + p_1 \overline{T}_{z,j} + p_0 \right) \nonumber \\
		&= n_{p,j} \Big( \tilde{p}_1 T_{z,j} + \tilde{p}_0 \Big). \label{eq:people_heat_rate_lin}
\end{align}
Recall now that the zone temperature profile $T_{z,j}$ to track is assumed to be linear in time. If we approximate the occupancy $n_{p,j}$ as a linear function of time within each time slot as well, as suggested in \cite{ISGT13}, then equation \eqref{eq:people_heat_rate_lin} can be analytically integrated from $(k-1)\dt$ to $k\dt$ to obtain the energy transferred to zone $j$ in the $k\th$ time slot:
	\begin{align*}
		E_{p,j}(k) =& \; q_{2,k}(n_{p,j})T_{z,j}(k\dt)  + q_{1,k}(n_{p,j})T_{z,j}((k-1)\dt) + q_{0,k}(n_{p,j})
	\end{align*}
where we set
\begin{equation}
	\begin{aligned}
		&q_{2,k}(n_{p,j}) = \dfrac{\tilde{p}_1 \dt}{6} \left( 2 n_{p,j}\left(k\dt\right) + n_{p,j}\left(\left(k-1\right)\dt\right)  \right) \\
		&q_{1,k}(n_{p,j}) = \dfrac{\tilde{p}_1 \dt}{6} \left( n_{p,j}\left(k\dt\right) + 2 n_{p,j}\left(\left(k-1\right)\dt\right)  \right) \\
		&q_{0,k}(n_{p,j}) = \dfrac{\tilde{p}_0 \dt}{2} \left( n_{p,j}\left(k\dt\right) + n_{p,j}\left(\left(k-1\right)\dt\right)  \right)
	\end{aligned}
	\label{eq:people_coeff}
\end{equation}

The total amount of energy transferred to all zones in each time slot can be packed in a vector $E_p(k) = [E_{p,1}(k) \cdots E_{p,n_z}(k)]\T$ and then, defining $\bs{E_p} = [E_p\T(1) \allowbreak \cdots \allowbreak E_p\T(M)]\T$ and $\bs{n_p} = [n_{p,1}(0)\, \allowbreak n_{p,1}(\dt) \allowbreak \cdots n_{p,1}(M\dt) \cdots n_{p,n_z}(0)\, \allowbreak n_{p,n_z}(\dt) \allowbreak \cdots n_{p,n_z}(M\dt)]$, one can write
that
\begin{equation}
	\bs{E_p} = N(\bs{n_{p}})\bs{u} + e(\bs{n_{p}}),
	\label{eq:people_energy_tot_matrix}
\end{equation}
where $N(\bs{n_{p}})$ and $e(\bs{n_{p}})$ depend on the coefficients~\eqref{eq:people_coeff}.

Note that occupancy profiles can be either obtained from data or derived from a stochastic model, \AF{like, e.g., the one in \cite{VBPSP2017} which is based on Poisson arrival/departure processes~\cite{harchol2013performance}.

Further energy contributions of the building occupants, in terms for instance of blinds movement and setpoint override, are not modeled here. Recent works on human-building interaction discuss the impact of human intervention on energy management strategies. The interested reader is referred to \cite{Burak2014}, where a possible strategy to limit human intervention is proposed, and to \cite{Gunay2014}, where a model predictive control solution is suggested for timely adjusting the control action to unpredicted human disturbances.}

\subsubsection{Other internal energy contributions}
There are many other types of heat sources that may affect the internal energy of a building, e.g., lighting, daylight radiation through windows, electrical equipment, etc. The overall heat flow rate produced within zone $j$ can be expressed as the sum of three contributions, namely
\begin{equation}
	Q_{\mathrm{int},j} = \alpha_j Q^S  + \lambda_j+ \kappa_j I_{\mathbb{R^+}}(n_{p,j}),
	\label{eq:windows_heat_rate}
\end{equation}
where $\alpha_j$ is a coefficient that takes into account the mean absorbance coefficient of zone $j$, the transmittance coefficients of the windows and their areas, sun view and shading factors, and radiation incidence angle. $I_{\mathbb{R^+}}(\cdot)$ denote the indicator function on the positive real values. The thermal energy contribution to zone $j$ due to internal lighting and electrical equipment is composed of two contribution: a constant term $\lambda_j$, and an additional therm $\kappa_j$ that represents the change in internal lighting and electrical equipment when people are present. Note that $Q_{\mathrm{int},j}$ does not depend on $Q^L$ because windows are usually shielded against longwave radiation.
\AF{The energy $E_{\mathrm{int},j}(k)$ during the $k^\mathrm{th}$ slot is given by:
\begin{align*}
E_{\mathrm{int},j}(k) =~& \dfrac{\dt}{2} \left[Q^S(k\dt) + Q^S((k-1)\dt)\right] + \dt\lambda_j\\
& +\dfrac{\dt}{2}
\kappa_j \left[ I_{\mathbb{R^+}}\left(n_{p,j}\left(k\dt\right)\right) + I_{\mathbb{R^+}}\left(n_{p,j}\left(\left(k-1\right)\dt\right)\right) \right]
\end{align*}
and is obtained by \eqref{eq:windows_heat_rate}, where the first (linear) and second (constant) terms have been analytically integrated, whereas the third term has been treated separately, due to the presence of the indicator function. In the cases when occupancy drops to zero or becomes nonzero in a time slot, the  energy contribution is set to a half of the contribution in the case when occupancy is nonzero at the beginning and at the end of the time slot.}
We can collect the thermal energy of the zones in a single vector $E_{\mathrm{int}}(k) = [E_{\mathrm{int},1}(k) \cdots E_{\mathrm{int},n_z}(k)]\T$, and  then define $\bs{E_\mathrm{int}} = [E_{\mathrm{int}}\T(1) \cdots E_{\mathrm{int}}\T(M)]\T$, which is finally given by:
\begin{equation}
\bs{E}_{\text{int}} = M \bs{\omega} + L(\bs{n_{p}}).
\label{eq:int_energy_tot_matrix}
\end{equation}

\subsubsection{Overall building cooling energy request}

Now we can finally compute the cooling energy demand of all zones in the building for tracking the piecewise linear zone temperature profiles $\bs{T_z}$ specified via the input $\bs{u}$ at the discrete time instants $k=0,1,\dots,M$ during the time horizon $[t_i,t_f]$.
Specifically, from \eqref{eq:cooling_energy_zone_j} it follows that $\bs{E_c}=[E_c\T(1) \cdots E_c\T(M)]\T$ with $E_c\T(k)=[E_{c,1}(k) \cdots E_{c,n_z}(k)]\T$ is the sum of four contributions:
\begin{align*}
\bs{E_c} = \bs{E}_{\bs{w}} + \bs{E}_{\bs{z}} + \bs{E}_{\bs{p}} + \bs{E}_{\text{int}},
\end{align*}
where $\bs{E}_{\bs{w}}$ is given in~\eqref{eq:wall_energy_tot_matrix}, $\bs{E}_{\bs{z}}$ in~\eqref{eq:zone_energy_equation}, $\bs{E}_{\bs{p}}$ in~\eqref{eq:people_energy_tot_matrix}, and $\bs{E}_{\text{int}}$ in~\eqref{eq:int_energy_tot_matrix}. This leads to the following expression for the cooling energy demand:
\begin{align*}
\bs{E_c} &= \tilde{F}\bs{T}(0) + (\tilde{G} + Z + N(\bs{n_{p}})) \bs{u} + (\tilde{H}+M)\bs{\omega}  + e(\bs{n_{p}})+ L(\bs{n_{p}})\\
&= \bs{A_c} \bs{T}(0) + \bs{B_c}(\bs{n_p}) \bs{u} + \bs{W_c}\bs{\omega} + \bs{b}(\bs{n_p})
\end{align*}
where $\bs{A_c}$, and $\bs{W_c}$ are constant matrices, whereas $\bs{B_c}(\bs{n_p})$ and $\bs{b}(\bs{n_p})$ depend on the occupancy.
Note that the input $\bs{u}$ defining the zone temperature profiles enters affinely the system dynamics if the occupancy $\bs{n_p}$ were fixed.

\subsubsection{Building block: interfaces and related constraints}

The thermal model of the building can be considered as a block with the following input/output interfaces:
the control input vector $\bs{u}$ specifying the piecewise linear  zone temperature profiles $\bs{T_z}$ at the discrete time instants $k=0,1,\dots,M$, and
disturbance input vectors $\bs{n_p}$ and $\bs{\omega}$  representing the occupancy and the collection of outdoor temperature $T_{out}$ and incoming shortwave $Q^S$ and longwave $Q^L$ radiations, respectively; and the output vector $\bs{E_c}$ of the cooling energy demand requested by the zones in the building to track $\bs{T_z}$.

Notice that the cooling energy demand cannot be negative. Furthermore, a profile where the zone temperature is required to decrease with a steep slope cannot be tracked. This can be formulated as a constraint  on the maximum amount of energy $E_{c,j}^{\max}$ that can be requested by a zone $j$ per time slot (from which the upper bounding vector $ \bs{E_{c}^{\max}}$ of the same size of $\bs{E_{c}}$ can be derived), and, possibly, a maximum amount $E_{c,b}^{\max}$ that can be requested by the building during the whole time horizon. This maps into the following actuation constraints:
\begin{align}\label{eq:building_bounds}
&0\leq \bs{E_{c}}\leq \bs{E_{c}^{\max}}, \quad \bs{1}\T \bs{E_c}\leq E_{c,b}^{\max},
\end{align}
where $\bs{1}$ denotes a column vector with all elements equal to $1$ so that $\bs{1}\T \bs{E_c}$ is the total cooling energy requested by the building. Note that when a vector is compared with a scalar like in \eqref{eq:building_bounds}, it means that each component of the vector is compared with that same scalar.

Table~\ref{Tab:summaryBuilding} summarizes the relevant quantities related to the building model. The \emph{type} attribute is introduced to denote possible different models that can be used, which eventually has some impact on the energy management problem formulation. Type A is the controllable building model where the zone temperature profiles can be optimized via the control input $\bs{u}$, whereas Type B is the uncontrollable building model where the zone temperature profiles cannot be chosen but are already specified via some given $\overline{\bs{u}}$ vector. In a network configuration, it is possible to include both controllable and uncontrollable buildings. Comfort and cooling energy bounds can then be enforced only in the case of Type A model, which contributes  to the network description with equations and inequalities that are linear in the control input.

\begin{table}
\linespread{1}
\centering
\resizebox{\textwidth}{!}{
\begin{tabular}{ccc|>{\centering\arraybackslash}p{1cm}|>{\centering\arraybackslash}p{1cm}|}
\cline{4-5}
& & & \multicolumn{2}{ c| }{\textbf{Model type}} \\ \cline{4-5}
& & & A & B\\ \hline
\multicolumn{1}{|c}{\multirow{2}{*}{\rotatebox[origin=c]{90}{\small\textbf{Model}}}} & \multicolumn{1}{|c|}{$\bs{E_{c}}=\bs{A_c} \bs{T}(0) + \bs{B_c}(\bs{n_p}) \bs{u} + \bs{W_c}\bs{\omega} + \bs{b}(\bs{n_p})$} & Linear in the control input & $\checkmark$ & -- \\ \cline{2-5}
\multicolumn{1}{|c}{}& \multicolumn{1}{|c|}{$\bs{E_{c}}=\bs{A_c} \bs{T}(0) + \bs{B_c}(\bs{n_p}) \overline{\bs{u}} + \bs{W_c}\bs{\omega} + \bs{b}(\bs{n_p})$} & Uncontrollable & -- & $\checkmark$ \\ \hline\hline
\multicolumn{1}{|c}{\multirow{4}{*}{\rotatebox[origin=c]{90}{\small\textbf{Variables}}}} & \multicolumn{1}{ |c| }{$\bs{u} \in \mathbb{R}^{n_z M}$}    & Control input & $\checkmark$ & -- \\ \cline{2-5}
\multicolumn{1}{|c}{}& \multicolumn{1}{ |c| }{$\bs{\omega} \in \mathbb{R}^{4 M}$} & Disturbance input   & $\checkmark$ & $\checkmark$ \\ \cline{2-5}
\multicolumn{1}{|c}{}& \multicolumn{1}{ |c| }{$\bs{n_p} \in \mathbb{R}^{n_z M}$}       & Disturbance   input & $\checkmark$ & $\checkmark$ \\ \cline{2-5}
\multicolumn{1}{|c}{}& \multicolumn{1}{ |c| }{$\bs{E_c} \in \mathbb{R}^{n_z}$}    & Output        & $\checkmark$ & $\checkmark$ \\ \hline\hline
\multicolumn{1}{|c}{\multirow{2}{*}{\rotatebox[origin=c]{90}{\small\textbf{Constr.}}}} &  \multicolumn{1}{ |c| }{$0\leq \bs{E_{c}}\leq \bs{E_{c}^{\max}}$}    & \multirow{2}{*}{Actuation} & $\checkmark$ & --  \\ \cline{2-2}\cline{4-5}
\multicolumn{1}{|c}{}& \multicolumn{1}{ |c| }{$\bs{1}\T \bs{E_c}\leq E_{c,b}^{\max}$}     &  & $\checkmark$ &  --\\ \hline
\end{tabular}
}
\caption{Summary of main characteristics of the building thermal model.}
\label{Tab:summaryBuilding}
\end{table}

\subsection{Chiller plant}\label{sec:model_chiller}
A chiller plant is an electrical devices that reduces the temperature of a liquid, typically water, via vapor compression or absorption cycle. In this way, it
converts the electric power provided by the electrical grid into cooling power, which is then conveyed to either some cooling load or some thermal storage via the chilled water circuit.

\begin{figure}[htb]
\centering
\includegraphics[scale=1]{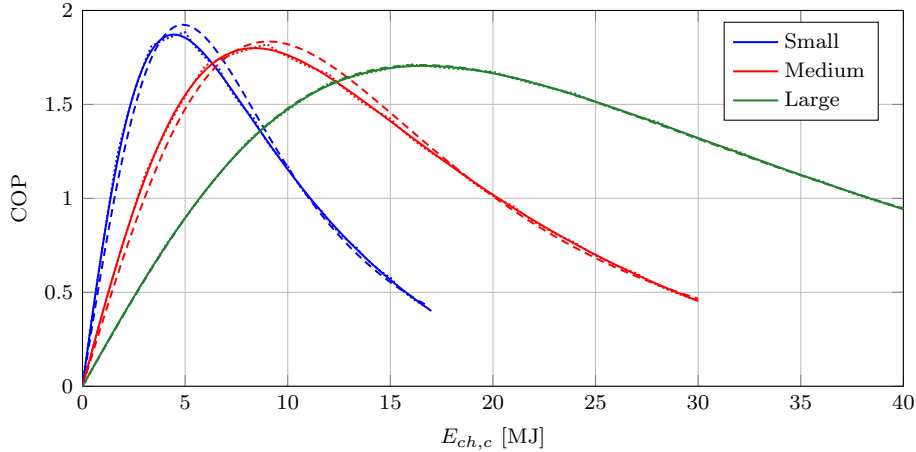}
\caption{COP curves for chillers of different size (solid lines), with their biquadratic approximations (dashed line), and their PWA approximations with $10$ knots (dotted line).}
\label{fig:chillerCOP}
\end{figure}

Chillers can be modeled through the equation
\begin{equation}\label{eq:chiller}
 E_{\ch,\ell}=\frac{a_1T_oT_{cw}\dt+a_2(T_o-T_{cw})\dt+a_4T_oE_{\ch,c}}{T_{cw}-\frac{a_3}{\dt}E_{\ch,c}}-E_{\ch,c}, \ 0\le E_{\ch,c}\le E_{\ch,c}^{\max},
\end{equation}
where $E_{\ch,\ell}$ is the electrical energy absorbed by the chiller in order to provide the cooling energy $E_{\ch,c}$  in a time slot of duration $\dt$, and $E_{\ch,c}^{\max}$ is the corresponding maximum cooling energy production.
Note that $E_{\ch,\ell}$ depends also on the outdoor temperature $T_o$ and the temperature of the cooling water $T_{cw}$. The latter is typically regulated by low level controllers so that it is maintained almost constant at some prescribed optimal operational value, which also facilitates the stratification in the thermal storage.
The chiller description \eqref{eq:chiller} is derived from the original Ng-Gordon model \cite{GordonNg} which is based on entropy and energy balance equations \AF{and accounts also for heat losses and  pump contribution to the electric energy consumption ($E_{\ch,\ell}> 0$ when $E_{\ch,c}=0$).}.
Coefficients $a_1$, $a_2$, $a_3$, $a_4$ characterize the chiller performance. Depending on their values, we can have different efficiency curves as given by the Coefficient Of Performance (COP), which is the ratio between the produced cooling energy and the corresponding electrical energy consumption:
\begin{align*}
\mathrm{COP} = \dfrac{E_{ch,c}}{E_{ch,\ell}}.
\end{align*}
Figure~\ref{fig:chillerCOP} shows an example of curves of the COP for three chiller units of different size, with their respective approximations presented in the following sections.

\AF{We next introduce simpler approximations of relation~\eqref{eq:chiller}, which preserve convexity in the control input $E_{\ch,c}$. }

\subsubsection{Chiller model approximations}
A convex biquadratic approximation
\begin{equation}
E_{\ch,\ell}=c_1(T_o)E_{\ch,c}^4+c_2(T_o)E_{\ch,c}^2+c_3(T_o), \ 0\le E_{\ch,c}\le E_{\ch,c}^{\max},
\label{eq:chiller_biquadratic}
\end{equation}
of the nonlinear Ng-Gordon model~\eqref{eq:chiller} can be derived by using weighted least square to best fit the most relevant points, i.e, those that correspond to zero energy request and to the maximum COP values.

Another possible convex approximation of \eqref{eq:chiller} is via a PieceWise Affine (PWA) function given by the following convex envelope of a finite number of affine terms
\begin{equation}
	E_{\ch,\ell}=\max\{m_{c}(T_o)E_{\ch,c}+q_{c}(T_o)\}, \ 0\le E_{\ch,c}\le E_{\ch,c}^{\max},
	\label{eq:chiller_pwa}
\end{equation}
where  the coefficients of the affine terms are collected in the two vectors $m_{c}(T_o)$ and $q_{c}(T_o)$, and the max operator is applied component-wise.
\AF{Note that, if $E_{\ch,\ell}$ in expression \eqref{eq:chiller_pwa} is to be minimized, then \eqref{eq:chiller_pwa} can be easily translated as a set of linear constraints with an epigraphic reformulation.}

\begin{figure}[t]
 \centering
 \includegraphics[width=\textwidth]{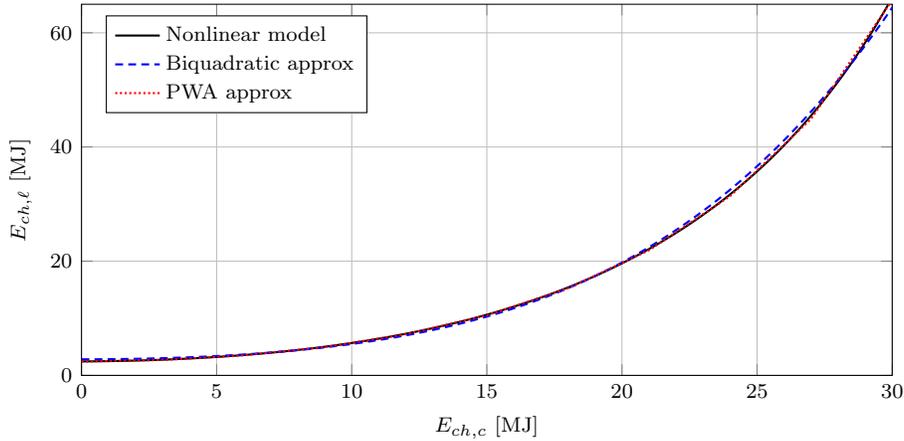}
 \caption{Simpler convex approximations of the electrical energy consumption as a function of the cooling energy request for the medium-size chiller unit.}
 \label{fig:chiller}
\end{figure}

The quality of the biquadratic and PWA approximations is compared in Figure \ref{fig:chiller}.

\subsubsection{On-off switching}
As shown in Figure \ref{fig:chiller}, the chiller absorbs some amount of electrical energy even when no cooling energy is produced.
In order to have the possibility of switching the chiller on and off, one can introduce the binary variable $\delta_\ch(k)$, \AF{$k=0,\dots,M$}, that represents the \emph{on} ($\delta_\ch(k)=1$) and \emph{off} ($\delta_\ch(k)=0$) logical status of the chiller at time $k$, $k=0,\dots,M$. The cooling energy request $E_{\ch,c}(k)$ and on-off command $\delta_\ch(k)$ are related via the logical condition
\begin{equation}\label{eq:on-off_ch}
 \delta_\ch(k) = 1 \quad \Leftrightarrow \quad E_{\ch,c}(k) > 0.
\end{equation}
Let $E_{\ch,c}^{\max}$ be the maximum value for $E_{\ch,c}$ and $\varepsilon$ a small quantity, typically set equal to the machine precision.
Using the \textit{Conjunctive Normal Form} in \cite{BM99},~\eqref{eq:on-off_ch} can be expressed as a mixed integer linear condition:
\[
\varepsilon \delta_\ch(k) \leq E_{\ch,c}(k) \leq E_{\ch,c}^{\max} \delta_\ch(k),
\]
which leads to $\delta_\ch(k) = 0 \Leftrightarrow E_{\ch,c}(k) =0$ and $\delta_\ch(k) = 1 \Leftrightarrow E_{\ch,c}(k) \in [\varepsilon,E_{\ch,c}^{\max}]$, that are practically equivalent to \eqref{eq:on-off_ch}.
Depending on the adopted approximation, we can rewrite the model of the chiller including the on-off condition as
\begin{align*}
 E_{\ch,\ell}(k)=
 \begin{cases}
 \left(c_1(T_o(k))E_{\ch,c}(k)^4+c_2(T_o(k))E_{\ch,c}(k)^2+c_3(T_o(k))\right)\delta_\ch(k)\\
 \max\{m_{c}(T_o(k))E_{\ch,c}(k)+q_{c}(T_o(k))\}\delta_\ch(k)
\end{cases}
\end{align*}
with $0\le E_{\ch,c}(k)\le E_{\ch,c}^{\max}$.
The PWA formulation is particularly convenient since the product between a (piecewise) affine function $Mx+q$ and a discrete variable $\delta$ can be reduced to a mixed integer linear condition \cite{BM99}, by introducing the auxiliary variable $ z=\delta (M x+Q)$ subject to $0\le z \leq \min\{M x+Q+(1-\delta) \BigM, \, \delta \BigM\}$, where $\BigM$ is an upper bound on $Mx+q$.

\subsubsection{Chiller block: interfaces and related constraints}

The chiller block can be described with a static map between the cooling energy  $\bs{E}_{\ch,c} = \left[E_{\ch,c}(0) \cdots E_{\ch,c}(M)\right]\T$ that it produces and the corresponding absorbed electrical energy $\bs{E}_{\ch,\ell} = \left[E_{\ch,\ell}(0) \cdots E_{\ch,\ell}(M)\right]\T$.

The cooling energy that the chiller can provide is subject to the following bound:
\begin{align*}
0\leq \bs{E}_{\ch,c}\leq E_{\ch,c}^{\max},
\end{align*}
which maps into a bound on the absorbed electrical energy
\begin{align*}
0\leq \bs{E}_{\ch,\ell}\leq E_{\ch,\ell}^{\max}.
\end{align*}

When the on-off command $\bs{\delta}_{\ch} = [\delta_{\ch}(0) \cdots \delta_{\ch}(M)]\T$ is introduced as an additional control input, the following further constraint enters the chiller model:
\begin{align*}
&\varepsilon \bs{\delta}_\ch \leq \bs{E}_{\ch,c} \leq E_{\ch,c}^{\max}\bs{\delta}_\ch.
\end{align*}

Table~\ref{Tab:summaryChiller} summarizes the relevant quantities of the chiller model, with Type A, B, C, and D representing possible modeling variants. The $\max$ operator is applied element-wise, and the symbol $*$ is the element-wise multiplication.

\begin{table}
\linespread{1}
\centering
\resizebox{\textwidth}{!}{
\begin{tabular}{ccc|>{\centering\arraybackslash}p{0.8cm}|>{\centering\arraybackslash}p{0.8cm}|>{\centering\arraybackslash}p{0.8cm}|>{\centering\arraybackslash}p{0.8cm}|}
\cline{4-7}
& & & \multicolumn{4}{ c| }{\textbf{Model type}} \\ \cline{4-7}
& & & A & B & C & D\\ \hline
\multicolumn{1}{|c}{\multirow{4}{*}{\rotatebox[origin=c]{90}{\small\textbf{Model}}}} & \multicolumn{1}{|c|}{$\bs{E}_{\ch,\ell}=c_1\bs{E}_{\ch,c}^4+c_2\bs{E}_{\ch,c}^2+c_3$} & Biquadratic& $\checkmark$ & -- & -- & -- \\ \cline{2-7}
\multicolumn{1}{|c}{} & \multicolumn{1}{|c|}{$\bs{E}_{\ch,\ell}=\max\{m_c\bs{E}_{\ch,c}+q_c\}$} & PWA & -- & $\checkmark$ & -- & -- \\ \cline{2-7}
\multicolumn{1}{|c}{} & \multicolumn{1}{|c|}{$\bs{E}_{\ch,\ell}=(c_1\bs{E}_{\ch,c}^4+c_2\bs{E}_{\ch,c}^2+c_3)*\bs{\delta}_\ch$} & Biquadratic with on-off & -- & -- & $\checkmark$ & -- \\ \cline{2-7}
\multicolumn{1}{|c}{} & \multicolumn{1}{|c|}{$\bs{E}_{\ch,\ell}=\max\{m_c\bs{E}_{\ch,c}+q_c\}*\bs{\delta}_\ch$} & PWA with on-off & -- & -- & -- & $\checkmark$ \\ \hline\hline
\multicolumn{1}{|c}{\multirow{3}{*}{\rotatebox[origin=c]{90}{\small\textbf{Variables}}}} & \multicolumn{1}{ |c| }{$\bs{E}_{\ch,c} \in \mathbb{R}^M$}    & Control input & $\checkmark$ & $\checkmark$ & $\checkmark$ & $\checkmark$ \\ \cline{2-7}
\multicolumn{1}{|c}{} & \multicolumn{1}{ |c| }{$\bs{\delta}_\ch \in \{0,1\}^M$}   & Control input & -- & -- & $\checkmark$ & $\checkmark$ \\ \cline{2-7}
\multicolumn{1}{|c}{} & \multicolumn{1}{ |c| }{$\bs{E}_{\ch,\ell} \in \mathbb{R}^M$}    & Output & $\checkmark$ & $\checkmark$ & $\checkmark$ & $\checkmark$\\ \hline\hline
\multicolumn{1}{|c}{\multirow{4}{*}{\rotatebox[origin=c]{90}{\small\textbf{Constraints}}}} & \multicolumn{1}{ |c| }{$0\leq \bs{E}_{\ch,\ell}\leq E_{\ch,\ell}^{\max}$}    & Electrical energy bounds & $\checkmark$ & $\checkmark$ & $\checkmark$ & $\checkmark$ \\ \cline{2-7}
\multicolumn{1}{|c}{} & \multicolumn{1}{ |c| }{$0\leq \bs{E}_{\ch,c}\leq E_{\ch,c}^{\max}$}    & Cooling energy bounds & $\checkmark$ & $\checkmark$ & $\checkmark$ & $\checkmark$ \\ \cline{2-7}
\multicolumn{1}{|c}{} & \multicolumn{1}{ |c| }{$\bs{E}_{\ch,c} \geq \varepsilon\bs{\delta}_\ch$}    & \multirow{2}{*}{Logical on-off condition} & -- & -- & $\checkmark$ & $\checkmark$ \\ \cline{2-2}\cline{4-7}
\multicolumn{1}{|c}{} & \multicolumn{1}{ |c| }{$\bs{E}_{\ch,c} \leq E_{\ch,c}^{\max}\bs{\delta}_\ch$}                 &  & -- & -- & $\checkmark$ & $\checkmark$ \\ \hline
\end{tabular}
}
\caption{Summary of the main characteristics of the chiller model.}
\label{Tab:summaryChiller}
\end{table}

 
\subsection{Storage}\label{sec:model_storage}
Thermal Energy Storages (TESs) are becoming widely used in medium size grids. TESs represent the most effective way, or even sometimes the only way, to take advantage of renewable energy sources. This is indeed the case for thermal solar energy and geothermal energy systems. In a smart grid context, they can be used as energy buffers for unbinding energy production from energy consumption. More specifically, in a district cooling scenario, a TES for cooling energy can shift the request of cooling energy production to off-peak hours of electrical energy consumption, make chillers operate in high-efficiency conditions, and smooth peaks of electrical energy request with benefits both for power production and distribution network systems, see e.g.~\cite{Deng,Kody,Borrelli2009,ISGT13}.

There are many different technical solutions to store thermal energy, the most widely used are fluid tanks and Phase Changing Materials (PCMs) storages.
\AF{We focus next on fluid tanks modeling, and add a note on how the model can be extended to PCMs storages in Remark \ref{rem:batteries}.}
From an energy management perspective we will make use of a black box model, derived based on system identification techniques, that uses the energy exchange (added or removed) as input and the thermal energy stored as output. The simplest model is a first order AutoRegressive eXogenous  (ARX) system
\begin{align}\label{eq:storage_one_step}
S(k+1) = aS(k) - s(k),
\end{align}
where $S(k)$ is the amount of cooling energy stored and $s(k)$ is the cooling energy exchanged ($s(k)>0$ if the storage is discharged, and  $s(k)<0$ if it is charged) in the $k\th$ time slot,  while $a \in (0,1)$ is a coefficient introduced to model energy losses.
\\
By unrolling the thermal storage dynamics in \eqref{eq:storage_one_step} we can express the cooling energy stored along the look-ahead discretized time horizon $[t_i,t_f]$ in a compact form as
\begin{equation}
\label{eq:storage}
\bs{S} = \Xi_0 S(0) + \Xi_1 \bs{s},
\end{equation}
where we set $\bs{S} = [S(1) \cdots S(M)]\T$, $\bs{s} = [s(0) \cdots s(M-1)]\T$, and $\Xi_0$ and $\Xi_1$ are suitably defined matrices.

A more sophisticated model can be obtained by introducing dissipation effects through the efficiency coefficients $\beta_C \in [0,1]$ and $\beta_D \in [0,1]$ for the charge/discharge dynamics as follows:
\begin{align}
 S(k+1)= aS(k)-\Bigl((1-\beta_C)\delta_C+(1+\beta_B)\delta_D\Bigr)s(k),
\label{eq:storageDetailed}
\end{align}
where $\delta_C(k) \in \{0,1\}$ and $\delta_D(k) \in \{0,1\}$ indicate the mode in which the storage is operated: $\delta_C(k) = 1$ and $\delta_D(k) = 0$, the storage is charged ($s(k)<0$), $\delta_C(k) = 0$ and $\delta_D(k) = 1$ the storage is discharged ($s(k)>0$), and $\delta_C(k) = \delta_D(k) = 0$ the storage is not used.
Notice that $\delta_C(k)$ and $\delta_D(k)$ are mutually exclusive, which can be coded via the constraint
\begin{align}
\delta_D(k) + \delta_C(k) \leq 1.
\label{eq:storageDetailed_c1}
\end{align}
It is possible to set minimum and maximum thresholds for the energy exchange rate in both the charging and discharging phases by constraining $s(k)$  as follows:
\begin{align}
\delta_D(k) s_{D}^{\min} + \delta_C(k) s_C^{\max} \leq s(k) \leq \delta_D(k) s_{D}^{\max} + \delta_C(k) s^{\min}_C
\label{eq:storageDetailed_c2}
\end{align}
with $s_{C}^{\max} < s_{C}^{\min}\leq 0$ and $0 \leq s_{D}^{\min} < s_{D}^{\max}$.
Note that if $\delta_C(k) = \delta_D(k) = 0$ (storage not in use), inequalities \eqref{eq:storageDetailed_c2} degenerate to the condition $s(k)=0$.

Model \eqref{eq:storageDetailed} is bilinear in the control inputs since  $\delta_C(k)$ and $\delta_D(k)$ are multiplied by $s(k)$. However, we can reduce it to the linear model
\begin{align}
 S(k+1)= aS(k)-(1-\beta_C)s_C(k)-(1+\beta_D)s_D(k)
\label{eq:storageDetailedReformulated}
\end{align}
by replacing $s(k)$ with the new control variables $s_C(k) = \delta_C(k)s(k)$ and $s_D(k) = \delta_D(k)s(k)$.
Accordingly, the constraint \eqref{eq:storageDetailed_c2} becomes
\begin{align}
&\delta_C(k)s_C^{\max} \leq s_C(k) \leq \delta_C(k) s_C^{\min}\label{eq:storageDetailedReformulated_c3}\\
&\delta_D(k)s_D^{\min} \leq s_D(k) \leq \delta_D(k) s_D^{\max}\label{eq:storageDetailedReformulated_c4}.
\end{align}
The energy exchange $s(k)$ can then be recovered from $s_C(k)$ and $s_D(k)$ as $s(k) = s_C(k)+s_D(k)$.

Model \eqref{eq:storageDetailed} subject to constraints \eqref{eq:storageDetailed_c1} and \eqref{eq:storageDetailed_c2} is equivalent to model \eqref{eq:storageDetailedReformulated}
subject to constraints \eqref{eq:storageDetailed_c1}, \eqref{eq:storageDetailedReformulated_c3} and \eqref{eq:storageDetailedReformulated_c4}. This latter model has the advantage of being linear so that it can be expressed in compact form along the look-ahead discretized time horizon $[t_i,t_f]$ as follows:
\begin{align*}
&\bs{S} = \Xi_0 S(0)+ \Xi_C \bs{s_C} + \Xi_D \bs{s_D}\\
&\bs{s} = \bs{s_D}+ \bs{s_C},
\end{align*}
where  $\bs{s_C} = [s_C(0) \cdots s_C(M-1)]\T$, $\bs{s_D} = [s_D(0) \cdots s_D(M-1)]\T$, and $\Xi_C$ and $\Xi_D$ are suitably defined matrices.
Note that those elements of vectors $\bs{s_C}$ and $\bs{s_D}$ that correspond to a zero charge and discharge command in vectors $\bs{\sigma_C} = [\sigma_C(0) \cdots \sigma_C(M-1)]\T$ and $\bs{\sigma_D} = [\sigma_D(0) \cdots \sigma_D(M-1)]\T$ are set to zero (see \eqref{eq:storageDetailedReformulated_c3} and \eqref{eq:storageDetailedReformulated_c4}.
Given that the charge and discharge commands are mutually exclusive, we have that  $\bs{s_C}\T\bs{s_D}=0$.

\begin{remark}[passive thermal storage]
The described thermal storage system is \emph{active} in that it can be directly operated by charge/discharge commands.
\emph{Passive} thermal storages are instead physical elements, like the walls of a building, that can accumulate and release thermal energy but are not directly charged or discharged. Even though it is more difficult in principle to take advantage of passive thermal storages since there is no direct way to control them, in Section~\ref{sec:Example1} we shall show how an optimal energy management strategy can exploit them.
\end{remark}

\begin{remark}[electric batteries and PCMs thermal storages]\label{rem:batteries}
Note that batteries for electrical energy storage can in principle be modeled in the same way~\cite{en6010444}. However, charging/discharging efficiencies depend on the battery State Of Charge (SOC) and energy losses can be related to the exchanged energy (exchange efficiency), so that a more complex model has to be specifically introduced. Also, additional constraints as for example the minimum and maximum charging time should be added to obtain a feasible operation of the battery.
\AF{PCMs thermal storages can be modeled as electric batteries with the fraction of liquid in the storage playing the role of the SOC in determining the model coefficients}.
\end{remark}

\subsubsection{Storage block: interfaces and related constraints}

The proposed model for the thermal storage has as control input the energy exchange $\bs{s}$, eventually decomposed into the charge and discharge inputs $\bs{s_C}$  and $\bs{s_D}$ activated by the mutually exclusive commands $\bs{\sigma_C}$  and $\bs{\sigma_D}$. The stored energy $\bs{S}$ is the output of the model in  both cases. Since the storage capacity is limited and the stored energy is a positive quantity, the following constraints apply
\begin{align*}
0\leq \bs{S} \leq {S}^{\max}.
\end{align*}
In addition, the amount of energy that can be exchanged per time unit is limited, and it cannot exceed certain thresholds, i.e., the bounds
\begin{align*}
{s}^{\min}\leq \bs{s} \leq {s}^{\max}
\end{align*}
apply to the energy exchange $\bs{s}$, or bounds \eqref{eq:storageDetailedReformulated_c3} and \eqref{eq:storageDetailedReformulated_c4} apply to the charge and discharge inputs $\bs{s_C}$  and $\bs{s_D}$.

Table~\ref{Tab:summaryStorage} summarizes the relevant quantities related to the storage model. 
\AF{The type attribute denotes possible different models for the storage.}
\begin{table}
\linespread{1}
\centering
\resizebox{\textwidth}{!}{
\begin{tabular}{ccc|>{\centering\arraybackslash}p{0.8cm}|>{\centering\arraybackslash}p{0.8cm}|}
\cline{4-5}
& & & \multicolumn{2}{ c| }{\textbf{Model type}} \\ \cline{4-5}
& & & A & B\\ \hline
\multicolumn{1}{|c}{\multirow{2}{*}{\rotatebox[origin=c]{90}{\scriptsize\textbf{Model}}}} & \multicolumn{1}{|c|}{$\bs{S} = \Xi_0 S(0) + \Xi_1 \bs{s}$} & Linear & $\checkmark$ & \\ \cline{2-5}
\multicolumn{1}{|c}{} & \multicolumn{1}{|c|}{$\bs{S} = \Xi_0 S(0) + \Xi_D \bs{s_D}+ \Xi_C \bs{s_C}$} & Linear with dissipation effects &  & $\checkmark$ \\ \hline\hline
\multicolumn{1}{|c}{\multirow{6}{*}{\rotatebox[origin=c]{90}{\scriptsize\textbf{Variables}}}} & \multicolumn{1}{ |c| }{$\bs{s}\in \mathbb{R}^M$}   & Control input & $\checkmark$ & \\ \cline{2-5}
\multicolumn{1}{|c}{} & \multicolumn{1}{ |c| }{$\bs{s_D}\in \mathbb{R}^M_{\geq 0}$}   & Control input & & $\checkmark$\\ \cline{2-5}
\multicolumn{1}{|c}{} & \multicolumn{1}{ |c| }{$\bs{s_C}\in \mathbb{R}^M_{\leq 0}$}   & Control input& & $\checkmark$\\ \cline{2-5}
\multicolumn{1}{|c}{} & \multicolumn{1}{ |c| }{$\bs{\delta_D}\in \{0,1\}^M$}   & Control input & & $\checkmark$\\ \cline{2-5}
\multicolumn{1}{|c}{} & \multicolumn{1}{ |c| }{$\bs{\delta_C}\in \{0,1\}^M$}   & Control input & & $\checkmark$\\ \cline{2-5}
\multicolumn{1}{|c}{} & \multicolumn{1}{ |c| }{$\bs{S} \in \mathbb{R}^M$}    & Output & $\checkmark$ & $\checkmark$\\ \hline\hline
\multicolumn{1}{|c}{\multirow{5}{*}{\rotatebox[origin=c]{90}{\scriptsize\textbf{Constraints$\,$}}}} & \multicolumn{1}{ |c| }{${s}^{\min}\leq \bs{s} \leq {s}^{\max}$}   & Energy exchange rate bounds & $\checkmark$ &\\ \cline{2-5}
\multicolumn{1}{|c}{} & \multicolumn{1}{ |c| }{$\bs{\delta_D} {s_D}^{\min}\leq \bs{s_D} \leq \bs{\delta_D}{s_D}^{\max}$}   & Energy exchange rate bounds (discharge) &  & $\checkmark$\\ \cline{2-5}
\multicolumn{1}{|c}{} & \multicolumn{1}{ |c| }{$\bs{\delta_C} {s_C}^{\min}\leq \bs{s_C} \leq \bs{\delta_C}{s_C}^{\max}$}   & Energy exchange rate bounds (charge) &  & $\checkmark$\\ \cline{2-5}
\multicolumn{1}{|c}{} & \multicolumn{1}{ |c| }{$\bs{\delta_C} + \bs{\delta_D} \leq 1$}   & Logical constraint &  & $\checkmark$\\ \cline{2-5}
\multicolumn{1}{|c}{} & \multicolumn{1}{ |c| }{$0\leq \bs{S} \leq {S}^{\max}$}    & Stored energy bounds & $\checkmark$ & $\checkmark$\\ \hline
\end{tabular}
}
\caption{Summary of the main characteristics of the thermal storage model.}
\label{Tab:summaryStorage}
\end{table}

 
\subsection{Combined Heat and Power unit: Microturbine}\label{sec:model_chp}
A Combined Heat and Power (CHP) unit is a device that jointly produces electricity and heat power while consuming primal energy (fossil fuels or hydrogen) with the purpose of reducing the amount of energy wasted in the environment. In most cases one of these two products is a byproduct. For example, modern power plants recover waste heat and deliver it for district heating purposes.
CHPs with large capacity are becoming widely used and highly performing. At the same time a number of micro-CHP solutions are being developed, the most promising ones being microturbines and fuel cells that convert gas or hydrogen into heat and electricity. Combined Cooling, Heat and Power (CCHP) devices are also available that convert part of the produced heat into cooling energy.

\begin{figure}[t]
 \centering
 \includegraphics[width=\textwidth]{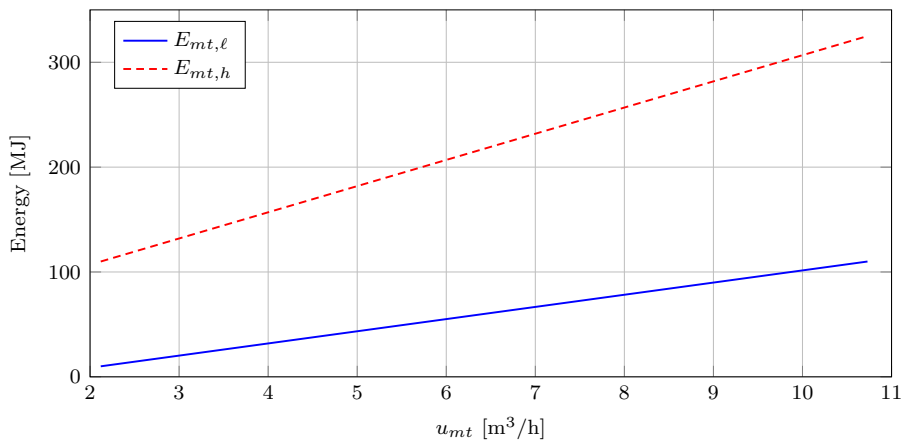}
 \caption{Characteristic curves of the C30 microturbine.}
 \label{fig:C30E}
\end{figure}

We consider a microturbine modeled through two static characteristics describing the electrical power production and the heat production, both as a function of the fuel volumetric flow rate. Figure~\ref{fig:C30E} represents the characteristics of the C30 microturbine produced by Capstone company \cite{C30TR}.
We can see that both curves are almost linear. The electrical energy $E_{\mt,\ell}(k)$ and the heat $E_{\mt,h}(k)$ produced by this microturbine during the $k\th$ time slot can then be expressed as affine functions of the fuel volumetric flow rate $u_\mt(k)$, that is supposed to be constant in each time slot, i.e.,
\begin{align*}
& E_{\mt,\ell}(k) = m_\ell u_\mt(k) + q_\ell,\\
& E_{\mt,h}(k) = m_h u_\mt(k) + q_h,
\end{align*}
where $m_\ell$, $q_\ell$, $m_h$, and $q_h$ are positive coefficients.

If we include the possibility of switching on or off the microturbine, we need to introduce the binary variable $\delta_\mt(k)$, $k=0,\ldots,M$, and modify the model as follows:
\begin{align*}
& E_{\mt,\ell}(k) = \delta_\mt(k)\left(m_\ell u_\mt(k) + q_\ell\right),\\
& E_{\mt,h}(k) = \delta_\mt(k)\left(m_h u_\mt(k) + q_h\right).
\end{align*}

Note that we do not model the microturbine transient from on to off, as instead suggested in~\cite{FGLSMA2002}. Yet, the static model that we adopt is accurate given that  a sensible choice of $\dt$ when addressing energy management is typically larger than the time scale of the microturbine dynamics.

\subsubsection{CHP block: interfaces and related constraints}

The CHP block represents a microturbine and is characterized by two control inputs that can be set in each time slot: the fuel volumetric flow rate $u_\mt$ and the on-off status of the microturbine $\delta_\mt$. It provides as outputs the electricity $E_{\mt,\ell}$ and the heat $E_{\mt,h}$ produced per time slot.

 Since the microturbine specifications require a minimum fuel volumetric flow rate $u_\mt^{\min}$ for the unit to be operative, we need to include the following logical condition:
\begin{equation*}
u_\mt(k) \le u_\mt^{\min} \quad \Leftrightarrow \quad \delta_\mt(k)=0,
\end{equation*}
which can be rewritten as:
\begin{align}
\delta_\mt(k)(u_\mt^{\min} +\varepsilon)\le u_\mt(k)\leq \delta_\mt(k)u_\mt^{\max}+(1-\delta_\mt(k)) u_\mt^{\min},
\label{eq:cph_min}
\end{align}
where $u_\mt^{\max}$ is the maximum flow rate and $\varepsilon>0$ is set equal to the machine precision.
The product between the (piecewise) affine function and a discrete variable $\delta$ is a nonlinear mixed integer expression that can be reduced to a mixed integer linear condition~\cite{BM99}.

The constraints related to the CHP are of three types:
\begin{enumerate}
\item Fuel inlet bounds of $\bs{u}_{\mt} = [u_\mt(0) \cdots u_\mt(M)]\T$:
\begin{align*}
u_\mt^{\min} \leq \bs{u}_\mt \leq u_\mt^{\max}
\end{align*}
\item Maximum heat and electrical energy that can be produced by the microturbine:
\begin{align*}
&0 \leq \bs{E}_{\mt,h} \leq E_{\mt,h}^{\max},\\
&0 \leq \bs{E}_{\mt,\ell} \leq E_{\mt,\ell}^{\max}
\end{align*}
with $\bs{E}_{\mt,h} = [E_{\mt,h}(0) \allowbreak\cdots \allowbreak E_{\mt,h}(M)]\T$, and $\bs{E}_{\mt,\ell} = [E_{\mt,\ell}(0)\allowbreak \cdots \allowbreak E_{\mt,\ell}(M)]\T$.
\item Logical on-off bounds~\eqref{eq:cph_min} expressed over the finite horizon $k=0,\ldots,M$ with $\bs{\delta}_{\mt} = [\delta_{\mt}(0) \cdots \delta_{\mt}(M)]\T$.
\end{enumerate}

Table~\ref{Tab:summaryCHP} summarizes the main characteristics of the CHP model.
\AF{Type A and B are the possible variants of the CHP model.}
\begin{table}
\linespread{1}
\centering
\resizebox{\textwidth}{!}{
\begin{tabular}{ccc|>{\centering\arraybackslash}p{1cm}|>{\centering\arraybackslash}p{1cm}|}
\cline{4-5}
& & & \multicolumn{2}{ c| }{\textbf{Model type}} \\ \cline{4-5}
& & & A & B\\ \hline
\multicolumn{1}{|c}{\multirow{4}{*}{\rotatebox[origin=c]{90}{\footnotesize\textbf{Model}}}} & \multicolumn{1}{|c|}{$\bs{E}_{\mt,\ell} = m_\ell \bs{u}_\mt + q_\ell$} & \multirow{2}{*}{Linear} & \multirow{2}{*}{$\checkmark$} & \multirow{2}{*}{--} \\
\multicolumn{1}{|c}{}  & \multicolumn{1}{|c|}{$\bs{E}_{\mt,h} = m_h \bs{u}_\mt + q_h$} &  &  & \\ \cline{2-5}
\multicolumn{1}{|c}{} & \multicolumn{1}{|c|}{$\bs{E}_{\mt,\ell} = (m_\ell \bs{u}_\mt + q_\ell) * \bs{\delta}_\mt$} & \multirow{2}{*}{Linear with on-off} & \multirow{2}{*}{--} & \multirow{2}{*}{$\checkmark$} \\
\multicolumn{1}{|c}{}  & \multicolumn{1}{|c|}{$\bs{E}_{\mt,h} = (m_h \bs{u}_\mt + q_h) * \bs{\delta}_\mt$} &  &  & \\ \hline\hline
\multicolumn{1}{|c}{\multirow{4}{*}{\rotatebox[origin=c]{90}{\footnotesize\textbf{Variables}}}}
 & \multicolumn{1}{ |c| }{$\bs{u}_{\mt} \in \mathbb{R}^M$}    & Control input & $\checkmark$ & $\checkmark$ \\ \cline{2-5}
\multicolumn{1}{|c}{} & \multicolumn{1}{ |c| }{$\bs{\delta}_\mt \in \{0,1\}^M$}    & Control input & -- & $\checkmark$ \\ \cline{2-5}
\multicolumn{1}{|c}{} & \multicolumn{1}{ |c| }{$\bs{E}_{\mt,\ell} \in \mathbb{R}^M$}    & Output & $\checkmark$ & $\checkmark$ \\ \cline{2-5}
\multicolumn{1}{|c}{} & \multicolumn{1}{ |c| }{$\bs{E}_{\mt,h} \in \mathbb{R}^M$}    & Output & $\checkmark$ & $\checkmark$
\\ \hline\hline
\multicolumn{1}{|c}{\multirow{3}{*}{\rotatebox[origin=c]{90}{\footnotesize\textbf{Constraints}}}} & \multicolumn{1}{ |c| }{$u_\mt^{\min}\leq \bs{u}_\mt \leq u_\mt^{\max}$}    & Fuel inlet bounds & $\checkmark$ & -- \\ \cline{2-5}
\multicolumn{1}{|c}{} & \multicolumn{1}{ |c| }{$\bs{u}_\mt \leq \bs{\delta}_\mt u_\mt^{\max}+(1-\bs{\delta}_\mt) u_\mt^{\min}$}    & \multirow{2}{*}{Logical on-off} & -- & $\checkmark$ \\ \cline{2-2}\cline{4-5}
\multicolumn{1}{|c}{} & \multicolumn{1}{ |c| }{$\bs{u}_\mt \geq (u_{\mt}^{\min} +\varepsilon) \bs{\delta}_\mt$}    &  & -- & $\checkmark$ \\ \hline
\end{tabular}
}
\caption{Summary of the main characteristics of the CHP model.}
\label{Tab:summaryCHP}
\end{table}

 
\subsection{Wind turbine}\label{sec:model_windturbine}

\begin{figure}[t]
 \centering
 \includegraphics[width=\textwidth]{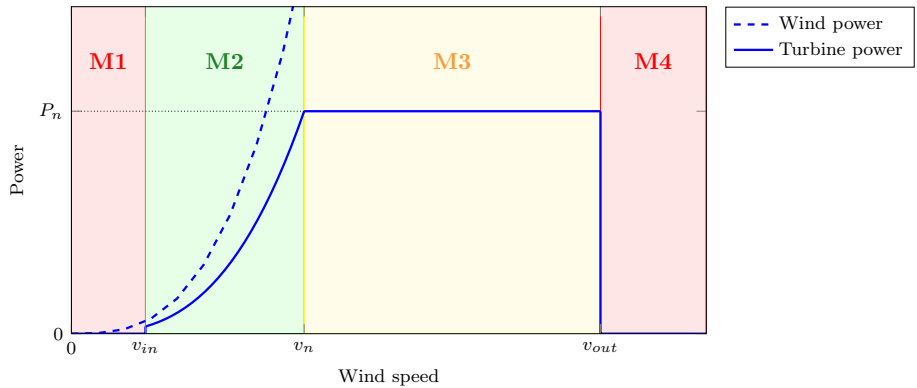}
 \caption{Characteristic curve of the power production by a wind turbine.}
 \label{fig:PowerProductionWT}
\end{figure}

A wind turbine is used to convert the wind kinetic energy into electrical energy.
Four different operational modes are typically defined for a controlled wind turbine (see  Figure \ref{fig:PowerProductionWT}): \emph{Mode 1}, when the wind speed value is within the range from zero up to the cut-in wind speed $v_{in}$ and there is no power produced by the wind turbine, which is turned off; \emph{Mode 2}, below the rated power $P_n$, thus called \emph{below-rated}, where the power captured from the wind is maximized; \emph{Mode 3}, above the rated wind speed, thus called \emph{above-rated}, where the wind turbine is saturated to the rated power $P_n$, and as the wind speed increases above the nominal turbine speed $v_{n}$, the blade pitch angle is adjusted so that local angles of attack acting on local airfoil sections become smaller, and hence the loads become relatively smaller and the power keeps constant; \emph{Mode 4}, when the wind speed is above the cut-out wind speed $v_{out}$, and the wind turbine is shut down, due to load and fatigue issues.
A turbine that is optimally sized for the site where it is installed is operating most of the time at the transition point between Mode 2 and Mode 3, also called \emph{at-rated}~\cite{Jonkman2009}. The power generated by the wind turbine $P_{wt}$ can then be calculated as follows:
\begin{align}
P_{wt} =
\begin{cases}
0,& v_{\text{wind}} \leq v_{in} \text{ or }  v_{\text{wind}} \geq v_{out}\\
P_m(v_{\text{wind}}),& v_{in}\leq v_{\text{wind}} \leq v_{n}\\
P_n,& v_{n}\leq v_{\text{wind}} \leq v_{out}
\end{cases}
\label{eq:windPowerModel}
\end{align}
where $P_{m}(v_{\text{wind}})$ is the maximum power that can be extracted from the wind kinetic energy when the wind speed is $v_{\text{wind}}$, while $P_{n}$ is the rated power.

Notice that the wind speed $v_{\text{wind}}$ is acting as a disturbance on the turbine. Therefore, the power produced by the wind turbine as output given the disturbance input $v_{\text{wind}}$ is a disturbance as well.
To the purpose of the energy management of the district network, we consider the static model in Figure~\ref{fig:PowerProductionWT} (solid line) for the power produced by the wind turbine as a function of the wind speed.  As for the wind speed prediction, both physical and statistical models, e.g., based on Markov chain, have been considered in the literature~\cite{Wu2007powertech,taylor2009tec,firat2010icmla,Papaefthymiou2008tec}. Combining~\eqref{eq:windPowerModel} with wind speed prediction models one can determine the energy contribution of the wind turbine by computing the average power produced within a time slot, and then multiplying it by the time slot duration $\dt$. 
\AF{Note that the static modeling of the wind turbine is appropriate if the time slot duration $\dt$ is sufficiently large compared to the involved inertia. In our set-up of a district network,  small scale wind turbines for roof installation could be included, compatibly with $\dt$ of the order of minutes.}

 \section{District network compositional modeling and optimal energy management}
\label{sec:configurations}

In this section, we show how the components previously introduced can be interconnected in order to define a certain district network configuration. 
\AF{We consider a network of buildings located in a neighborhood and do not model the distribution network.} 
Since the input/output interfaces of each component have been described in terms of thermal or electrical energy received or produced, energy balance equations and energy conversion functions can be adopted to combine the network components. For instance, the sum of the cooling energy requests of the buildings in the network should be equal to the sum of the cooling energy provided by chillers and taken from/stored in the thermal storages; each chiller receives as input a cooling energy request and provides as output the corresponding electrical energy consumption; the sum of the electrical energy consumption should be equal to the electrical energy produced by the local power generators, i.e. the CHP units and the wind turbine, taken from/stored in the batteries, and provided by the main grid.
Depending on the number of components and the adopted model for each component, the overall model of the district network has a different size and complexity, the most general one being hybrid due to the presence of both continuous and discrete variables, and stochastic due to the disturbances (e.g., occupancy, outside temperature, solar radiation, wind velocity) acting on the system, \cite{LygerosPrandini10}.

\begin{figure}[h]
 \centering
 \includegraphics[scale=0.5,keepaspectratio=true]{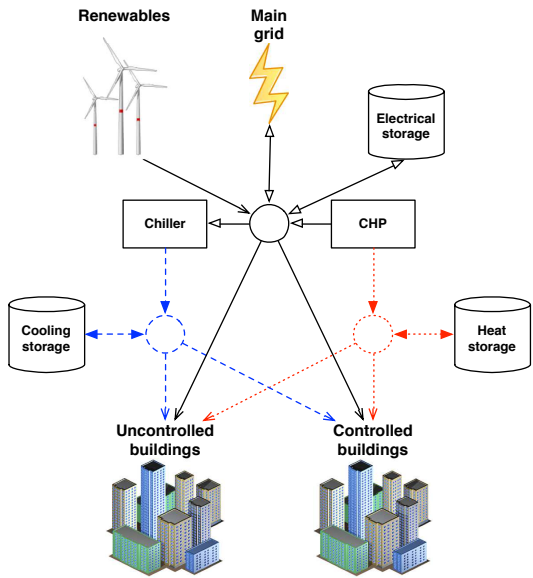}
 \caption{District network configuration.  The line style encodes the kind of energy: black thick, red thin, and blue dotted for electrical, heating, and cooling energy, respectively.
Different arrowheads are used for energy fluxes that can be controlled, controlled only indirectly, or not controlled. }
 \label{fig:Grid1}
\end{figure}
Figure~\ref{fig:Grid1} shows a possible district network configuration and the energy fluxes between its components and the main grid. The district network may be composed of multiple  buildings that share common resources such as cooling and heat storages, chillers, CHP units, batteries and renewable energy generators. The three nodes appearing in the figure do not correspond to any physical component but are introduced to point out that fluxes associated with the same kind of energy (electrical, heat, and cooling energy) add up to zero.
Some energy contributions can be controlled (e.g., those related to storage units), some others can be controlled only indirectly (e.g., electrical energy requested by the chiller), or cannot be controlled (e.g., renewable energy production). This is pointed out using different arrowheads in Figure~\ref{fig:Grid1}.
As for buildings, some of them are \emph{controlled} in that their cooling energy request can be modulated to some extent via the zone temperature setpoint. If the zone temperature setpoints are fixed and given by some comfort profiles, then the building is \emph{uncontrollable}.

\AF{We assume that the district network is connected to the main grid, which supplies the electrical energy needed to maintain the balance between electrical energy demand and generation within the district network.}

The district network is ``smart'' if it is possible to appropriately set the controllable variables so as to optimize its behavior.
A typical goal is to minimize the overall cost while guaranteeing the satisfaction of the energy needs of the users in the district.
Costs are mainly due to the electrical energy requested to the main grid and additional costs related to device operation such as startup and fuel costs. The overall cost is then given by:
\begin{equation}
 \label{cost}
J = C_\ell + C_\ch + C_\mt + C_f,
\end{equation}
where the first term is the electrical energy cost $C_\ell = \sum_{k=1}^M C_\ell(k)$; $C_\ch = \sum_{k=1}^M C_\ch(k)$ is the cost for the chillers startup; $C_\mt = \sum_{k=1}^M C_\mt(k)$ and $C_f=\sum_{k=1}^M C_f(k)$ are the costs for the CHPs startup and fuel consumption.

It is worth noticing that startup costs also serve the purpose of favoring solutions that avoid continuous and unrealistic switching of devices.
Note also that additional logical conditions are needed to account for them. For example, a chiller startup cost can be modeled as $C_\ch^\on  \max \{ \delta_\ch(k) - \delta_\ch(k-1), 0 \}$, where $C_\ch^\on$ is the actual startup cost which is accounted for at $k$ only if the chiller was off at $k-1$ and is switched on at $k$.
Similarly, for the CHP, its startup cost at $k$ is given by $C_\mt^\on  \max \{ \delta_\mt(k) - \delta_\mt(k-1), 0 \}$.
The fuel costs of a CHP are proportional to the amount of fuel consumption during the $k\th$ time slot, i.e., $\psi_f  \delta_\mt(k) u_\mt(k)  \dt$, where $\psi_f$ is the unitary fuel cost.

As for the electrical energy cost, the cost per time slot $C_\ell(k)$ is typically given by a PWA function of the electrical energy exchange $E_L(k)$ with the main grid, i.e.,
\begin{align}
\label{cost-electricity}
C_{\ell}(k)=\max\{c_{1,\ell}(k)E_L(k)+c_{0,\ell}(k)\},
\end{align}
where the coefficients of the affine terms are collected in vectors $c_{1,\ell}(k)$ and $c_{0,\ell}(k)$, and the max operator is applied componentwise. This expression allows us to adopt different values for revenues ($E_L(k)<0$) and actual costs ($E_L(k)>0$), and to account for penalties when the electrical energy consumption/production $E_L(k)$ exceeds certain thresholds. 
\AF{Note that, if $C_{\ell}$ is to be minimized, an epigraphic reformulation can be adopted to rewrite \eqref{cost-electricity} in terms of a set of linear inequalities.}

To describe $E_L$ for an arbitrary configuration, we adopt in this section the following short-hand notations. Components correspond to energy contributions and are defined through letters (building $\B$, chiller $\C$, storage $\St$, CHP microturbine $\CHP$) with a superscript that denotes the model type (symbols are given in Tables~\ref{Tab:summaryBuilding}--\ref{Tab:summaryCHP}) and the kind of energy (electrical $\ell$, cooling $c$, and heating $h$) provided as output. This is important, e.g., to distinguish between a thermal storage ($\St^c$) and an electric battery ($\St^\ell$), and also in the case when a component allows for multiple kinds of energy as output. For instance, $\CHP^{B,h}$ stands for the heating energy produced by a CHP described by a linear on-off model. The subscript possibly denotes the energy request received as input, as in the case of a chiller that has to provide the net cooling energy requested by buildings after deduction (addition) of that provided (requested) by the thermal storage units.

We can for example derive the expression of $E_L$  for the configuration in Figure~\ref{fig:Grid1}:
\begin{equation}
 \label{composition}
E_L= \C^{A,\ell}_{\leftarrow\{\B^{B,c} + \B^{A,c} + \St^c\}}+\CHP^{B,\ell}+\St^\ell. \end{equation}

If we then plug~\eqref{composition} into equation~\eqref{cost-electricity} and~\eqref{cost}, we get the expression for the cost function $J$ to be minimized.

Note that $J$ may be uncertain if there are disturbance inputs acting on the system. In such a case, one can either neglect uncertainty and refer to some nominal profile for the disturbance inputs or account for uncertainty and formulate a worst case or an average cost criterion based on $J$.
Furthermore, when we compose a district network model plugging together all the elements, we also get a number of constraints associated with them.
Constraints express both technical limits (i.e., maximum cooling energy that a chiller can provide) and performance requirements (i.e., comfort temperature range). Additional constraints can be added if needed (e.g., the maximum amount of electrical energy that the main grid can provide). Yet, constraints might be uncertain due to the presence of disturbances, and, hence they might be enforced only for the nominal profile, thus neglecting uncertainty, or as robust or probabilistic constraints.

Different approaches can then be adopted to address the energy management of the district network, depending also on the choice of the cost criterion (nominal/worst-case/average) and the constraints (nominal/robust/probabilistic).
\AF{Uncertainty on the parameters values could also be explicitly accounted for in the design. For instance, one could assume that parameters take equally likely values in some range and impose that performance is optimized over almost all instances except for a small set}.\\
Furthermore, different architectures (centralized, decentralized or distributed) can be conceived and implemented for the resulting optimization problem solution, depending on the actual communication and computation capabilities available in the network, and on possible privacy of information issues like in the case of a building that is not willing to share its own consumption profile, while still aiming at cooperating for reducing the overall district cost.

The formulation of the optimal energy management problem involves defining the following quantities:
\begin{enumerate}
\item \emph{Global parameters}, i.e., sampling time $\dt$, and number of $M$ of time slots of the look-ahead time-horizon. 

\item \emph{Optimization variables}, i.e., the decision variables to be set by the optimization problem. Notice that energy balances must always hold, and this may decrease the actual degrees of freedom of the system. For example, in the case of a controllable building with a chiller plant, the cooling energy request to the chiller cannot be set freely, since it has to match the cooling energy needed for the building to track the temperature setpoint that becomes effectively the only decision variable.

\item \emph{Cost function}, i.e., the quantity that has to be minimized, e.g., the electric energy costs or the deviation of the energy consumption from some nominal profile agreed with the main grid operator.

\item \emph{Constraints}, i.e., the feasibility conditions that limit the solution space of the optimization problem. Notice that constraints can be classified in three categories:
\begin{enumerate}
\item \emph{Single component constraints}, which are enforced at the level of each component separately and are related due to its dynamics and capabilities.
    For example, the energy accumulated in a storage is jointly dictated by the storage capacity and dynamics of the storage system.

\item \emph{Interconnection constraints}, which relate variables of different components and originate from their cooperative interaction in the district. For example, the temperature setpoint in a controllable building cannot result in a cooling energy request that is larger than the energy that the chiller can produce and the energy that can be taken from the storage.

\item \emph{Control constraints}, which are enforced to achieve some desired property of the energy management strategy. These are, for instance, the comfort constraints imposed on the temperature in a building or the constraints enforced at the end of the control time-horizon on the energy in the storage to avoid its depletion and allow for repetitive use of the control strategy in  a periodic fashion.
\end{enumerate}
\end{enumerate}

 \section{Some numerical examples}
\label{sec:simResults}

In this section, we present some examples where a district network configuration is considered and a related energy management problem is defined and solved.
\AF{All examples refer to a centralized architecture, with known profiles for the disturbances. 
We consider a one-day time horizon since this is a commonly used time horizon for building energy management, especially temperature control. 
We enforce a periodic solution to cope with the myopic attitude of the finite horizon strategy, which would empty the battery/storage and drive the zone temperatures to the limit of their admissible range at the end of the time horizon in order to save money, without caring of the next day.}

\AF{Examples were chosen to be simple but realistic enough to highlight the capabilities of the proposed framework.
Many more examples could be presented with reference to different set-ups in terms of either district network configuration or energy management problem formulation. 
Distributed energy management strategies could be adopted for easing computations and preserving privacy of information, as suggested in \cite{bfimgp2016}.  The stochastic nature of disturbances could be accounted for via a randomized approach as in \cite{ioli2016acc}, which however refers to a single building configuration. Stochastic periodic control solutions, \cite{FDIGP2017}, could be implemented as well. }

\subsection{Cooling of a controlled building with a chiller plant}
\label{sec:Example1}

\begin{figure}[h]
 \centering
 \includegraphics[scale=0.5,keepaspectratio=true]{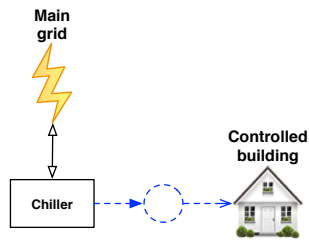}
 \caption{Configuration with a three story building connected to the chiller.}
 \label{fig:ThreeStory_configuration}
\end{figure}

Inspired by the numerical example presented in~\cite{ioli_adchem2015}, we start by considering the simple district network configuration in Figure~\ref{fig:ThreeStory_configuration}, which consists of a controlled building and a chiller unit.

The controlled building is a medium-sized three story office building of the following dimensions: $20$m long, $20$m wide, and $10$m tall. Each facade of the building is half glazed and the roof is flat. The biquadratic approximation~\eqref{eq:chiller_biquadratic} is used for the chiller model.

\begin{figure}[htb]
\centering
\includegraphics[width=\textwidth]{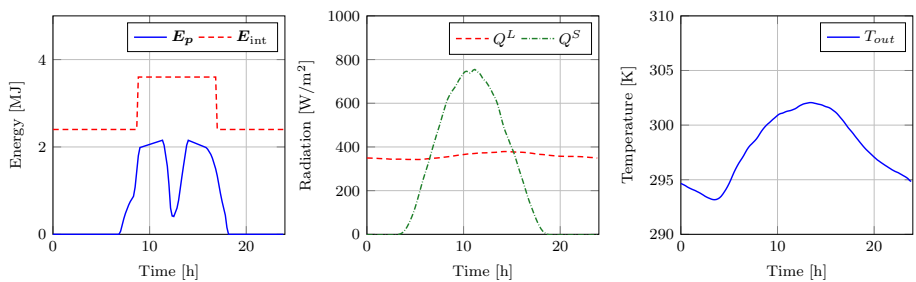}
\caption{Disturbances acting on the building: occupancy and internal energy contributions, solar radiation and outside temperature (from left to right).}
\label{fig:Example1_Disturbances}
\end{figure}

Disturbances are treated as deterministic signals. Figure~\ref{fig:Example1_Disturbances} shows the profiles adopted for the occupancy and internal energy contributions, solar radiation and outside temperature.

In Section \ref{sec:Example1}, we consider a single-zone setup for the building, where the three floors are treated as a unique thermal zone, with the same temperature setpoint. In Section \ref{sec:Example2}, we move to a multi-zone setup, where each floor is a thermal zone with its temperature setpoint.
\AF{In both cases we neglect energy exchanges through thermal radiation among internal walls and we consider the ground floor thermally isolated from the ground.}

\subsubsection{Single zone building}
\label{sec:Example1}

The purpose of this example is twofold:
\begin{enumerate}
\item Showing the role of the building structure as a passive thermal storage, that can accumulate and release thermal energy;
\item Compare the energy management strategies obtained with two different control objectives.
\end{enumerate}

The problem is then formulated as follows:
\begin{enumerate}
\item \emph{Global parameters}, the sampling time is set to $\dt = 10$ minutes, and the time horizon is set to 1 day, i.e., $M=144$.

\item \emph{Optimization variables}, the only optimization variable is the temperature set-point of the single zone $T_z$ as defined via the control input $\bs{u}$ over the considered finite horizon.

\item \emph{Cost function}, we here consider the two different cost functions:
\begin{enumerate}
\item cooling energy provided by the chiller:
\begin{align}
J_1 = \sum_{k=0}^M E_{\ch,c}(k).
\label{eq:J1_example}
\end{align}
\item electricity consumption:
\begin{align}
J_2 = \sum_{k=0}^M E_{\ch,\ell}(k),
\label{eq:J2_example}
\end{align}
\end{enumerate}
where the electricity consumption is related to the cooling energy of the chiller static characteristic that maps one into the other according to a specific COP (see Section~\ref{sec:model_chiller}).

\item \emph{Constraints}, the following constraints are included in the optimization problem:
\begin{enumerate}
\item \emph{Single component constraints}, the constraints of the single components are given by
\begin{align*}
& \bs{E_{c}} \geq 0\\
& 0\leq \bs{E}_{\ch,c}\leq E_{\ch,c}^{\max}.
\end{align*}

\item \emph{Interconnection constraints:} the chiller satisfies the cooling load demand, i.e.,
\begin{align*}
&E_{\ch,c}(k) = E_{c}(k), \quad k=0,\dots,M.
\end{align*}

\item \emph{Control constraints:} zone temperature should be within some comfort range, and a periodic solution is enforced by setting the same value for the zone temperature setpoint at the beginning and end of the time horizon:
\begin{align*}
& u^{\min}\le \bs u\le u^{\max}\\
&\bs u(M) = \bs u(0) =\bs{T_z}(0)\\
& \bs{T}(0) = \bs{T}(M)
\end{align*}
\end{enumerate}
We consider an ideal setting where both $\bs{T_z}(0)$ and $\bs{T}(0)$ can be set so as to obtain periodic solution.
\end{enumerate}

The resulting optimization problem is a convex constrained program that can be solved, for example, with CVX\footnote{\url{http://cvxr.com/}} with the SDPT3 solver.

\begin{figure}[h!]
 \centering
 \includegraphics[width=\textwidth]{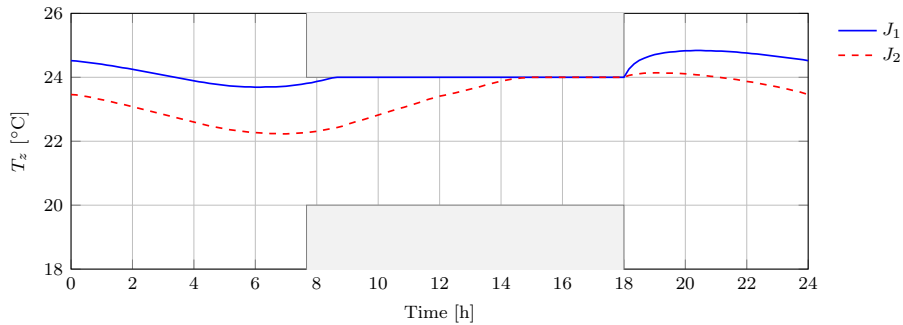}
 \caption{Temperature profiles obtained as solutions of the two optimization problems.}
 \label{fig:E1_Temperatures}
\end{figure}
\begin{figure}[h!]
 \centering
\includegraphics[width=\textwidth]{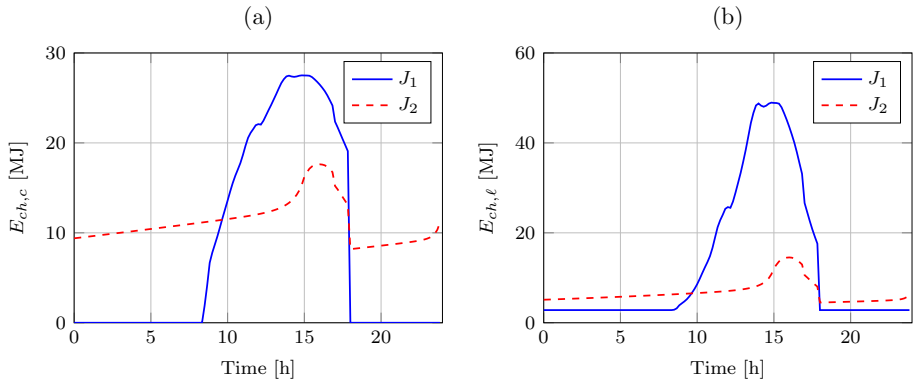}
 \caption{Cooling energy request (a) and absorbed electric energy (b) for the cost functions $J_1$ (cooling request) and $J_2$ (electric energy consumption).}
 \label{fig:E1_Energy}
\end{figure}

Figure~\ref{fig:E1_Temperatures} shows the resulting optimal temperature profiles $T_z$ for the two cases.
Both solutions stay within the prescribed comfort temperature bounds. Notice that the discrepancy between the two curves is at most of about $1.6^\circ$C. Despite this distance being small, from Figure~\ref{fig:E1_Energy} one can notice a clear difference in the required cooling energy for the two cases.
In the case of minimization of the electricity consumption ($J_2$), a ``precooling'' phase occurs from time 18:00 to time 10:00 of the next day (if we think about the solution applied over multiple days), which leads to a larger cooling energy request.

Intuitively, the second policy stores some cooling energy in the building structure, ahead of time, thus smoothing the cooling energy request in the central part of the day, when occupancy is larger, to get the chiller operating with high efficiency. The ``building thermal mass'' is  exploited as a passive thermal storage to add further flexibility to the system~\cite{Balaras96,Emery,Borrelli2012,ioli_adchem2015}.

\begin{figure}[h!]
 \centering
\includegraphics[width=\textwidth]{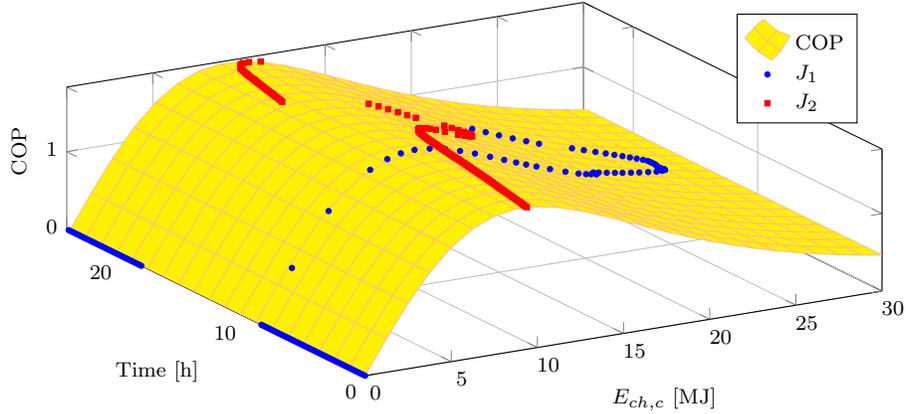}
 \caption{Values taken by the chiller COP in the two cases when the cooling energy request ($J_1$) and the electric energy consumption ($J_2$) are adopted as cost functions. }
 \label{fig:E1_COP}
\end{figure}

On the other hand, the first policy exploits the fact that at night the temperature is lower, comfort constraints are satisfied (indeed they are set to be larger because there are no occupants in the office building), and the chiller does not need to provide any cooling energy to the load. Figure~\ref{fig:E1_Energy} shows the electric energy consumption in the two cases, highlighting that the chiller is working at its minimum for most of the time in the cooling energy minimization policy ($J_1$). The integral of the curves in Figure~\ref{fig:E1_Energy} is the electricity consumption and is larger for the cooling energy minimization policy. Indeed, Figure~\ref{fig:E1_COP} shows that the second policy makes the chiller operate close to the maximum COP value of the chiller, thus requiring much less electrical energy.

In summary, depending on the cost function adopted in the energy management strategy design, one can have significantly different behaviors of the same district network configuration, with a different performance, even with a limited difference in the temperature setpoint profiles.

\subsubsection{Multi-zone building} \label{sec:Example2}

In this example, the controlled building has three thermal zones, one per floor, with three zone temperature setpoints, and our goal is to investigate the impact of a time-varying electricity price over the day time.

\begin{figure}[htb]
\centering
\includegraphics[width=\textwidth]{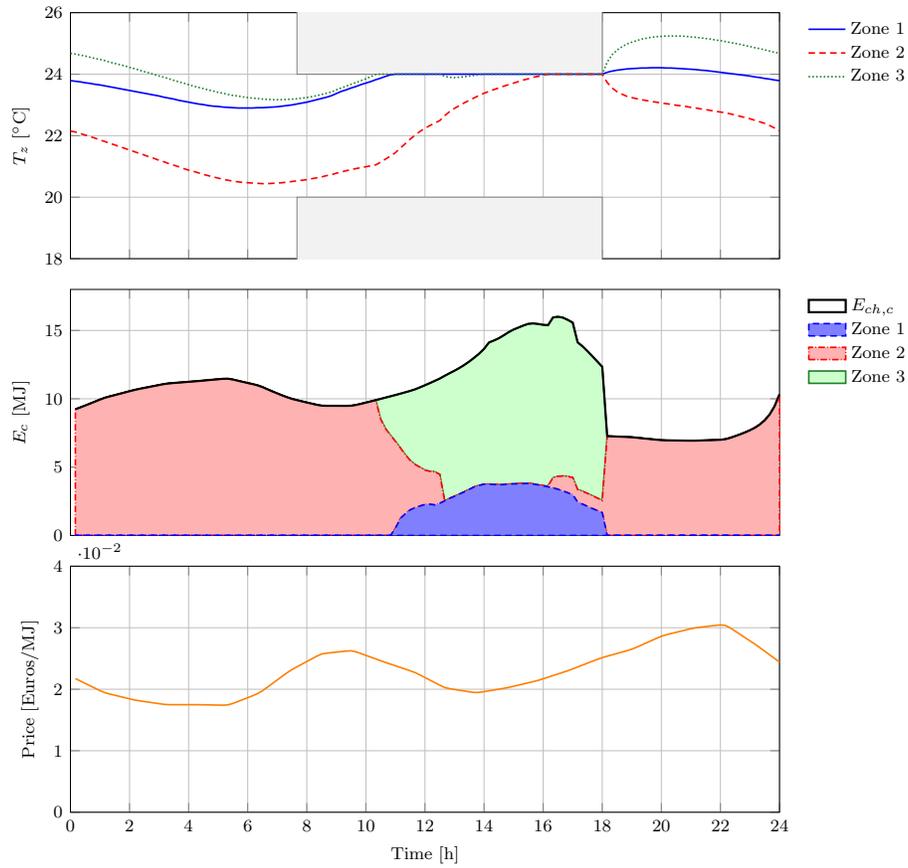}
\caption{Temperatures of the three zones (top graph) cooling energy profile of the three zones (middle graph) and price of the energy (bottom graph).}
\label{fig:Example2_Temperatures}
\end{figure}

The problem is formulated as follows:
\begin{enumerate}
\item \emph{Global parameters}, the sampling time is set to $\dt = 10$ minutes, and the time horizon is set to 1 day, i.e., $M=144$.

\item \emph{Optimization variables}, the optimization variables are the temperature setpoints of the three zones $\bs{T_z} = [T_{z,1}\;T_{z,2}\;T_{z,3}]\T$, as defined via the control input $\bs{u}$ over the considered finite horizon.

\item \emph{Cost function}, we minimize the cost of the electrical energy needed for cooling the building
\begin{align*}
J = \sum_{k=0}^M p(k) E_{\ell}(k),
\end{align*}
where $p(k)$ is the time-varying unitary cost (see Figure \ref{fig:Example2_Temperatures}).

\item \emph{Constraints}, the following constraints are included in the optimization problem:
\begin{enumerate}
\item \emph{Single component constraints}, the constraints of the single components are given by
\begin{align*}
& \bs{E_{c}} \geq 0\\
& 0\leq \bs{E}_{\ch,c}\leq E_{\ch,c}^{\max}.
\end{align*}

\item \emph{Interconnection constraints}, the chiller satisfies the cooling load demand, i.e.,
\begin{align*}
&E_{\ch,c}(k) = E_{c}(k), \quad k=0,\dots,M.
\end{align*}

\item \emph{Control constraints},
zone temperature should be within some comfort range, and a periodic solution is enforced by setting the same value for the zone temperature setpoint at the beginning and end of the time horizon:
\begin{align*}
& u^{\min}\le \bs u\le u^{\max}\\
&\bs u(M) = \bs u(0) =\bs{T_z}(0)
\end{align*}
\end{enumerate}
\end{enumerate}

Figure~\ref{fig:Example2_Temperatures} shows the optimal temperature setpoints for the three zones (top graph), the stack of the cooling energies associated with each zone and how they compose the cooling energy request $E_{\ch,c}$  to the chiller (middle graph), and the unitary price for the electrical energy. The obtained solution shows that all the zones are precooled in the first hours of the day. However, all the cooling energy provided by the chiller is conveyed to Zone 2. This can be explained by observing that Zone 2 is the central story, and it exchanges thermal energy with the other zones through its ceiling and floor, thus cooling them. In addition, considering jointly the cooling energy request profile with the price of energy, it is possible to see that whenever the price of energy is decreasing the energy request increases, and \emph{vice versa}. Finally, notice that at the end of the day, Zone 2 is being cooled again, to bring the temperature back to its value at the beginning of the day, with a larger amount of cooling energy request when the price gets lower.

\begin{figure}[htb]
\centering
\includegraphics[scale=1]{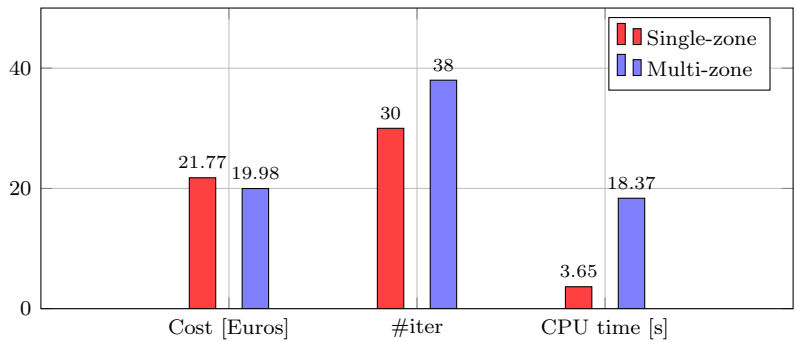}
\caption{Comparison between the single-zone and the multi-zone setting in terms of cost and number of iterations.}
\label{fig:Example2_bars}
\end{figure}

The example highlights the advantages of defining multiple thermal zones in a building. Figure~\ref{fig:Example2_bars} shows the total cost obtained for the single-zone and multi-zone cases, and the number of iterations needed to find the optimal solution in the two cases. The total cost decreases by $8\%$ with the multi-zone configuration, but the computational complexity increases as witnessed by the higher number of iterations of the solver and the total CPU time. This analysis can be performed at design time to decide the number of zones to be included, based on the available computational resources and the economical advantage.

\subsection{Microgrid with uncontrolled building}

\begin{figure}[h]
 \centering
 \includegraphics[scale=0.5,keepaspectratio=true]{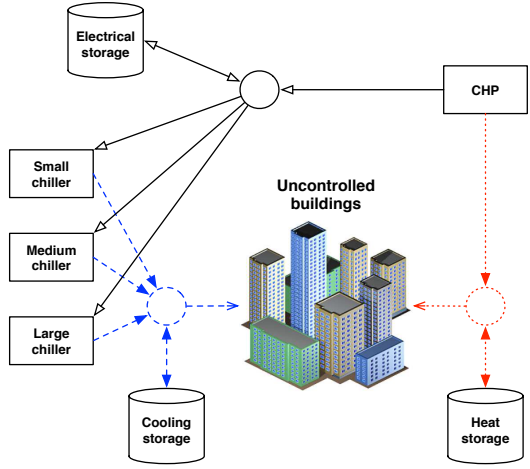}
 \caption{Microgrid configuration diagram.}
 \label{fig:E7_Dia}
\end{figure}

In this example, we consider a microgrid configuration as represented in Figure~\ref{fig:E7_Dia}, which includes a set of uncontrolled buildings, a cooling and a heat storage, three chillers of different sizes (small, medium, and large, indexed by 1, 2 and 3, respectively, and whose COP curves are presented in Figure~\ref{fig:chillerCOP}), a microturbine, and an electrical storage (a battery).

Buildings are modeled via their requests of cooling energy $E_{c}$ (for air temperature control) and heating energy $E_{h}$ (for getting warm water). The chillers and microturbine can be switched on and off by setting the associated logical variables  $\delta_{\ch,i}$, $i =1,2,3$ and $\delta_\mt$ to 1 and 0.
We adopt the PWA approximation~\eqref{eq:chiller_pwa} for the three chillers, with $10$ knots, obtained from their Ng-Gordon models with \AF{$T_o = 22^\circ$C, $T_{cw} = 10^\circ$C,} and the following parameters:
\begin{align*}
&\text{Chiller 1 (small):}~  & a_1=0.0056, \quad a_2=10.11,\quad a_3=7.00,\quad a_4=0.9327, \\
&\text{Chiller 2 (medium):}~ & a_1=0.0109, \quad a_2=20.22,\quad a_3=3.80,\quad a_4=0.9327, \\
&\text{Chiller 3 (large):}~  & a_1=0.0230, \quad a_2=40.44,\quad a_3=1.98,\quad a_4=0.9327.
\end{align*}
\AF{The measure units of the coefficients are $[a_1] = \frac{W}{K}$, $[a_2] = W$, $[a_3] = \frac{K}{W}$, $[a_4] = 1$.}

The problem is then formulated as follows:
\begin{enumerate}
\item \emph{Global parameters}, the sampling time is $\dt = 1$ hour, and the time horizon is 1 day, hence the number of time lots is $M=24$.

\item \emph{Optimization variables}, the optimization variables are the chillers and the microturbine on/off status $\delta_{\ch,i}$, $i= 1,2,3$, and $\delta_{\mt}$, the cooling energy requested to the chiller $E_{\ch,c}$, the CHP fuel inlet $u_\mt$, and the energy exchanged with the storage units $s_{i}$, $i \in \{c,h,\ell\}$.

\item \emph{Cost function}, composed of different components:
\begin{enumerate}
\item The cost for energy trading:
\begin{align*}
C_{\ell}(k) = p(k) \left(E_\ell(k) - E_{\mt,\ell}(k) -s_{\ell}(k)\right)
\end{align*}
where $p(k)$ is the unitary cost for trading electrical energy (the same used for the previous experiment), and $E_\ell(k)$ is the electrical energy demand;

\item The fuel cost for the microturbine given by:
\begin{align*}
C_{f}(k) = \psi_f(k)  \delta_\mt(k) u_\mt(k)
\end{align*}
where $\psi_f$ is the unitary cost for the fuel, here considered constant and unitary: $\psi(k) = 1$.

\item The startup costs of the microturbine:
\begin{align*}
C_{\mt}(k) = C_{\mt}^\on  \max \{ \delta_{\mt}(k) - \delta_{\mt}(k-1), 0 \}
\end{align*}
where $C_{\mt}^\on =1$.

\item The startup costs of the chillers:
\begin{align*}
C_{\ch}(k) = \sum_{i=1}^3 C_{\ch,i}^\on  \max \{ \delta_{\ch,i}(k) - \delta_{\ch,i}(k-1), 0 \}
\end{align*}
where $C_{\ch,1}^\on =0.05$, $C_{\ch,2}^\on =0.1$, $C_{\ch,3}^\on =0.2$.
\end{enumerate}
Therefore, the overall cost function becomes:
\begin{align*}
J = \sum_{k=0}^M C_{\ell}(k) + C_{f}(k) + C_{\mt}(k) + C_{\ch}(k)
\end{align*}

\item \emph{Definition of constraints}, the following constraints are included in the optimization problem:
\begin{enumerate}
\item \emph{Single component constraints} given by:
\begin{align*}
&\delta_\mt(k)(2 +\varepsilon)\leq u_\mt(k) \leq 10\delta_\mt(k)+ 2(1-\delta_\mt(k)) ,\\
&0 \leq E_{\mt,h}(k) \leq 2160\text{MJ}, \quad 0 \leq E_{\mt,\ell}(k) \leq 1080 \text{MJ},\\
&-500\text{MJ}\leq s_{h}(k) \leq 500 \text{MJ}, \quad
0\leq S_{h}(k) \leq 1500 \text{MJ},\\
&-360\text{MJ}\leq s_{c}(k) \leq 360 \text{MJ}, \quad 0\leq S_{c}(k) \leq 1800 \text{MJ},\\
&-250\text{MJ}\leq s_{\ell}(k) \leq 250 \text{MJ},\quad 0\leq S_{\ell}(k) \leq 1500 \text{MJ},\\
&\varepsilon \delta_{\ch,i}(k) \leq E_{\ch,i}(k) \leq  \delta_{\ch,i}(k) \cdot 180 \text{MJ}, \quad \forall i=\{1,2,3\}.
\end{align*}

\item \emph{Interconnection constraints}, given by thermal energy balance equations:
\begin{align*}
&\sum_{i=1}^{3} E_{\ch,c,i}(k) + s_c(k) = E_{c}(k),\\
&E_{\mt,h}(k) + s_h(k) = E_{h}(k)
\end{align*}

\item \emph{Control constraints}, a periodic solution is enforced by setting the same value for the on/off status of the devices and for the stored energies: \begin{align*}
&\delta_{\mt}(0) = \delta_{\mt}(M)\quad \delta_{\ch,i}(0) = \delta_{\ch,i}(M), \quad i = 1,2,3\\
& S_{h}(0)= S_{h}(M)\quad S_{c}(0)= S_{c}(M) \quad S_{\ell}(1)= S_{\ell}(M)
\end{align*}
\end{enumerate}
\end{enumerate}

The resulting optimization problem is a Mixed Integer Linear Programming (MILP) problem, and can be solved using YALMIP to formulate the optimization problem and CPLEX as a solver.

\begin{figure}[h!t]
\centering
\includegraphics[width=\textwidth]{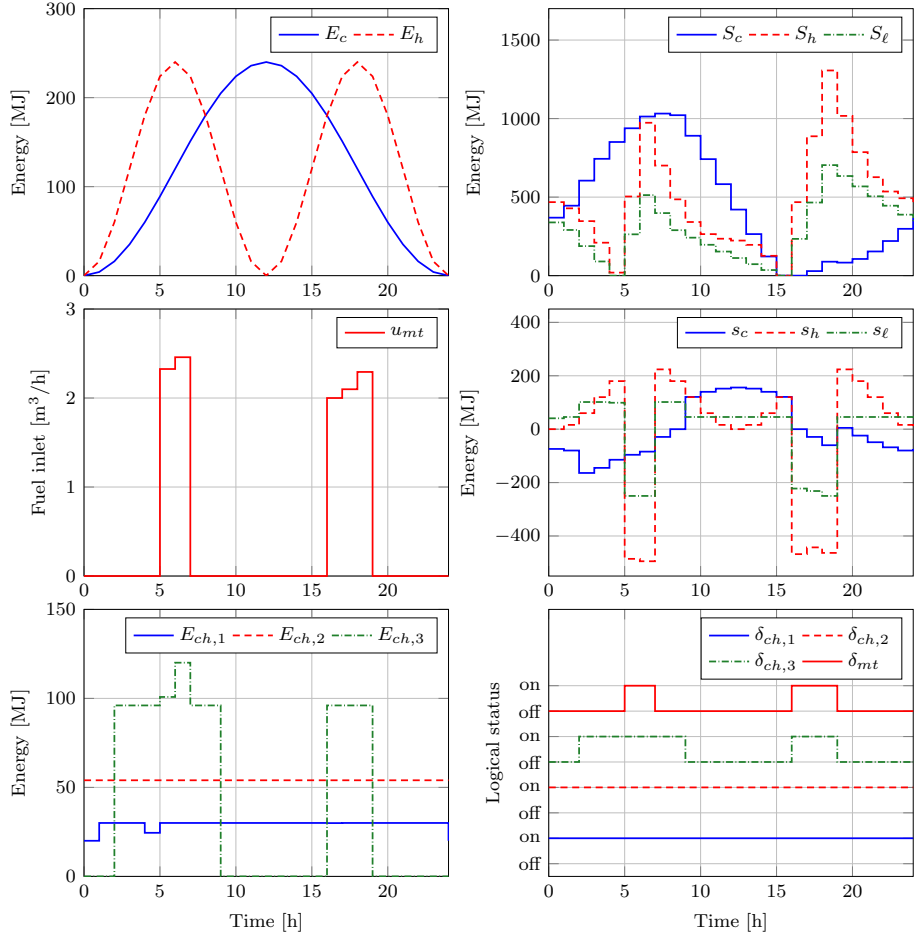}
\caption{Optimal energy management strategy: behavior of the relevant quantities.}
\label{fig:Example7_Energy}
\end{figure}

Figure~\ref{fig:Example7_Energy} shows the cooling energy demand $E_c$ and the heating energy demand $E_h$ of the uncontrolled buildings over the considered time horizon (top left graph), the storage of cooling, heat and electrical energy (top right graph), with the respective control inputs (middle right graph), the fuel inlet to the microturbine (middle left graph), the energy demand for the three chillers (bottom left graph), and finally, the bottom right graph shows the logical on/off status of the chillers and of the microturbine.

Notice that the small and medium size chillers (associated with $\delta_{\ch,1}$ and $\delta_{\ch,2}$) are never switched off, while the large chiller is used to charge the cooling storage before the peak of cooling energy demand, and right after that, when the storage got empty.
The microturbine is switched on only during the peaks of heat energy demand (see the profile of $\delta_{\mt}$).

\begin{figure}[htb]
\centering
\includegraphics[width=\textwidth]{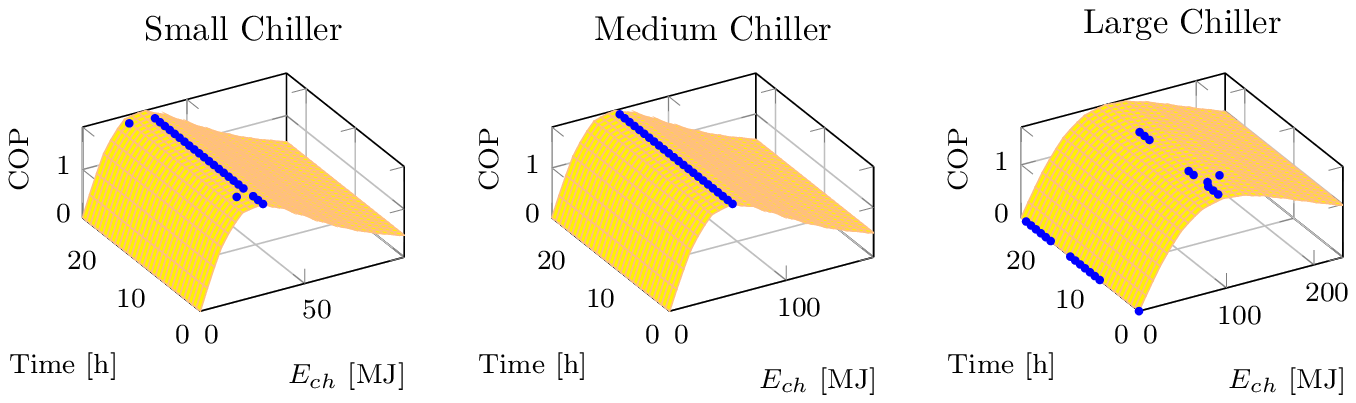}
\caption{Performance for the three chillers in terms of evolution in time of their COP.}
\label{fig:Example7_cop}
\end{figure}
Figure~\ref{fig:Example7_cop} shows the performance of the three chillers in terms of COP, as a function of time. It is possible to see that the small and medium chiller are always operating close to their maximum COP, while the large chiller either operates at its maximum COP or it is in the off mode, where its consumption is zero.

The cooling energy storage is charged at the beginning of the day when the cooling load is low, and then discharged when the load increases. It is finally recharged to meet the periodicity imposed via the control constraints.
On the other hand, the heating storage is used to meet the heating energy demand, and the microturbine is activated only when the heating storage is empty. Also in this case, part of the energy produced when the CHP was active, is stored to meet the energy demand of the next day.
 \section{Multirate control}
\label{sec:multirate}

Increasing the number buildings and/or thermal zones per building necessarily leads to a larger computational effort for solving the energy management control problem since the number of optimization variables increases as well.
This may become an issue when a receding horizon strategy is adopted and optimization is performed online at every control instant. Indeed, real-time constraints can hamper the applicability of the approach.

A possible way to avoid this issue is to use larger values of the sampling time $\dt$ for the discretization of the model, thus yielding a lower number of optimization variables for the same time horizon. As a side (positive) effect, discretizing with a larger sampling time make less stringent
the real time constraints since it gives more time to perform the computations and apply the solution.

Unfortunately, using a larger sampling time degrades the model accuracy, thus eventually deteriorating the control performance.
This issue can be tackled by taking a \emph{multirate control approach}, where model and controller operate with different sampling periods. Specifically, if we let $\dt$ be the sampling period of the model and introduce the rate $M_R\in\mathbb{N}$, then, in multirate control, the control action is only set every $M_R$ time slots of length $\dt$, or, equivalently, $\du = M_R \dt$ is the sampling period for the controller.
This choice allows for an accurate representation of the model dynamics, while still decreasing the number of optimization variables, and, as a consequence, the computational complexity, by a factor $M_R$. 
Clearly, the reduction of the number of optimization variables has some impact on the achievable performance in terms of cost and also reactiveness to possible disturbances with fast dynamics. The choice of the rate $M_R$ must compromise between computational effort reduction and performance degradation, compatibly with the available resources.

\subsection{Example}
Let us focus on the example presented in Section~\ref{sec:Example1}, with cost function given by the electrical energy
\begin{align*}
J = \sum_{k=0}^M E_{\ell}(k).
\end{align*}
We sample the model with $\dt = 10$ minutes, and we study the effects of employing different rates $M_R$ for applying the control input, namely $M_R = 1,6,12,24,36,48$, corresponding to $\du = \frac{1}{6},1,2,4,6,8$ hours, thus progressively reducing the number of optimization variables.

\begin{figure}[h]
 \centering
 \includegraphics[width=\textwidth]{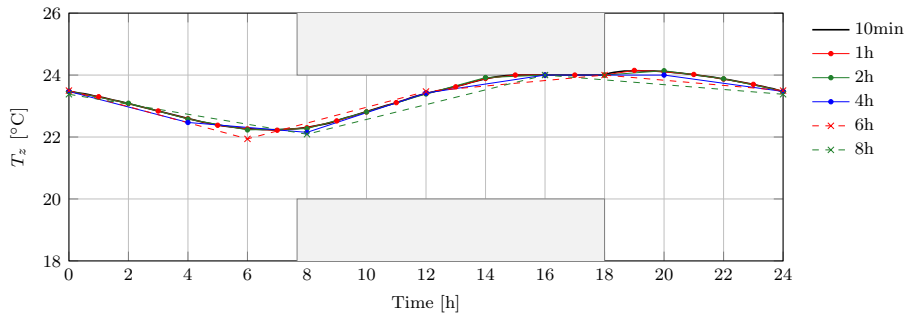}
 \caption{Optimal temperature profile obtained with different control rates.}
 \label{fig:E3_Temperatures}
\end{figure}

\begin{figure}[h]
 \centering
\includegraphics[width=\textwidth]{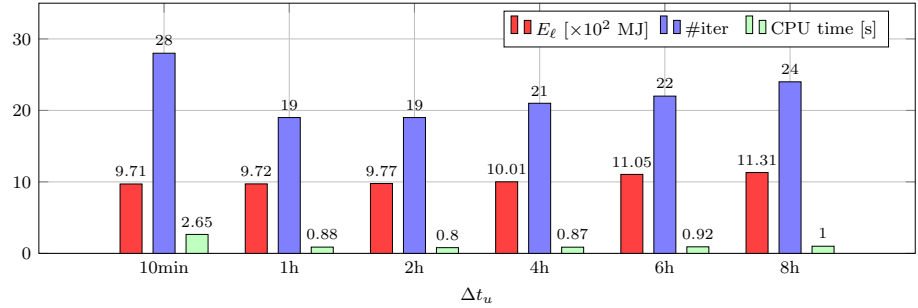}
 \caption{Performances evaluation with different control rates.}
 \label{fig:E3_JMultirate}
\end{figure}

Figure~\ref{fig:E3_Temperatures} shows the optimal temperature profiles for the different rates. Notice that the curves associated with $\du = 10$ minutes ($M_R = 1$) and with $\du = 1$ hour ($M_R = 6$) are practically indistinguishable, but in the latter case we have reduced the number of optimization variables by a factor 6. 
The reduction of the optimization variables causes an increase of the costs, as shown in Figure~\ref{fig:E3_JMultirate}. However, this increase is negligible up to $\du = 2$ hours, while the computation effort is almost constant for values of $\du$ larger than or equal to 1 hour, if evaluated in terms of total CPU time\footnote{The total CPU time presented in Figure~\ref{fig:E3_JMultirate} is computed as the average total CPU time over $100$ experiments for each considered $\du$, for the sole solver.}.

\begin{figure}[h]
 \centering
 \includegraphics[width=\textwidth]{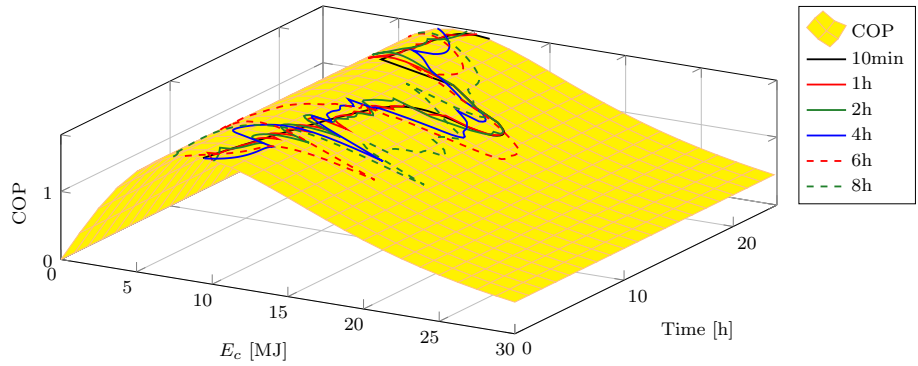}
 \caption{Performance for the chiller with different control rates.}
 \label{fig:E3_COP}
\end{figure}

We can also analyze the chiller performance presented in Figure~\ref{fig:E3_COP}.
The higher the rate $M_R$, the lower is the flexibility of the control input to do a fine adjustment of the temperature setpoint and compensate the disturbance variability. This is why the chillers are not constantly operating at a high efficiency levels when $M_R$ is larger. This results in a less performing chiller, so one should look for a tradeoff among computational effort and cost efficiency.

Finally, we can conclude that the adoption of multirate control solutions is problem specific, depending on the available computational power, and on the complexity of the optimization problem to be solved. 
 \section{Conclusion}
\label{sec:conclusion}

In this paper we presented a  modeling framework for the optimal operation of a district network, with reference in particular to the cooling of multiple buildings that are sharing resources like chillers or storages. Various components have been introduced and modeled in terms of energy fluxes so as to ease their composition via energy balance equations. A control-oriented perspective is adopted in that control and disturbance inputs are explicitly accounted for, in terms of their energy contribution.
We also described how to formulate an optimal energy management problem as a constrained optimization program where control inputs are the optimization variables  and need to be set so as to minimize some energy-related function (e.g., electric energy cost, deviation from some nominal profile of electric energy consumption), while satisfying comfort and actuation constraints.
Finally, a multirate approach was proposed to reduce the number of optimization variables while preserving the model accuracy. This has potential for real time applicability of the method when implemented according to the receding horizon strategy of model predictive control.
\AF{This will allow to compensate for unpredictable human-building interactions as discussed in \cite{Gunay2014}. }

Some numerical examples were also presented to show the versatility of the proposed framework.
Currently, we are addressing optimal energy management of a district network in presence of stochastic disturbances, the key challenge being how to account for them when embedded in a distributed setting with limited communications capabilities. \AF{The approach in \cite{mfgp2017} could be useful to this purpose.}

\appendix
\section{Model validation}
\label{sec:validation}

Reliability of the model is crucial when adopting model-based control design strategies. At the same time, if a model is accurate but very complex, then, design can become impractical.

As for what concerns the network district modeling for energy management purposes, the more difficult component to model is the building, since various factors need to be accounted for, including size and structure of the building, walls composition, presence of electrical devices, occupancy, and environmental conditions, like outdoor temperature and solar radiation. Also, model complexity grows as the size of the system increases.

Models and modeling frameworks for buildings have been proposed in the literature~\cite{wetter2014jbps,bonvini2011mcmds,crawley2000energyplus,fumo2010methodology,Kontes2014846}. Most of them include a detailed characterization of the fluid dynamics phenomena, e.g., the evolution of the temperature and humidity of the thermal zones, and they typically require specialized Computational Fluid Dynamics (CFD) tools for simulation. Even though these approaches provide very accurate simulation results, they are difficult to use for control design purposes, due to their complexity. In this paper we adopted a control-oriented perspective and presented a simple model of the building where thermal zone temperatures act as control inputs and enters linearly the system dynamics.

Validation of a model of the building dynamics against experimental data is quite challenging, also because setting up a measurement facility for a building can be complex and expensive. In order to validate the presented model, we hence resort to the methodology introduced by the American Society for Heating Refrigerating and Air-conditioning Engineers (ASHRAE), and, more specifically, the validation method defined in the ANSI-ASHRAE 140 standard.
This standard specifies test cases and procedures for the evaluation of the technical capabilities and range of applicability of computer programs that compute the thermal performance of buildings and their HVAC systems. The current set of tests included in the standard consists of (i) comparative tests that focus on building thermal envelope and fabric loads, and mechanical equipment performance, and (ii) analytical verification tests that focus on mechanical equipment performance. Different building energy simulation programs, with different levels of modeling complexity, can be tested. For all tests included in the specifications, results provided by other certified simulation tools are presented, and they represent the baseline for validating new modeling and simulation software. A detailed description of the simulation tools included in the specification can be found in~\cite{bonvini2012phdthesis}.

We here provide the results of some of the main tests defined in the ANSI-ASHRAE 140 standard, and compare them with the baseline provided in the standard. Let us first introduce the test case and then describe the validation procedure.

We consider a building located at an altitude of $1649$m above the sea level, and weather data series resuming the weather conditions for a whole year are available and provided by the standard. The data set contains: external dry bulb temperature, wind speed, wind direction, direct and diffuse solar radiation.
\begin{figure}
\centering
\includegraphics[scale=1]{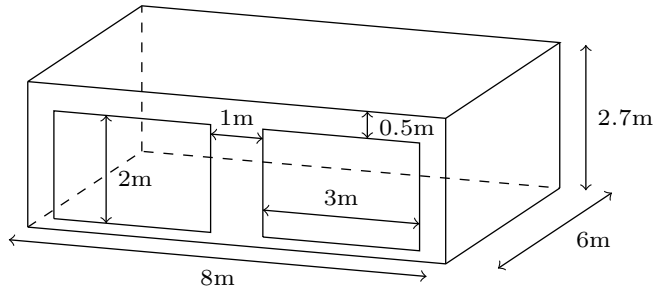}
\caption{Room geometry with isometric South windows.}
\label{fig:buildingScheme}
\end{figure}

The building has a $48$m$^2$ floor area, a single story with rectangular-prism geometry, and two south-facing windows, $6$m$^2$ each (see Figure~\ref{fig:buildingScheme}). Two set-up are considered, which differs in materials composition and walls thickness: lightweight (case 600 in the standard) and heavyweight (case 900 in the standard). The standard specifies the composition in terms of thickness, density, thermal conductivity and specific heat capacity of all layers of each wall, for both the lightweight and heavyweight cases. These values are listed in Tables~\ref{tab:material_lightweight} and~\ref{tab:material_heavyweight}, respectively. According to the specification, density and specific heat of the underfloor insulation have the minimum admissible value allowed in the building model that is tested, and in any case a value not smaller  than zero. Also, the contribution of the internal loads and people, within the thermal zone is constant over the year and equal to $ Q_{\text{int}} + Q_p = 200$W.

\begin{table}[htb]
\centering
\caption{Walls composition for the lightweight building case. }
\label{tab:material_lightweight}
\resizebox{\textwidth}{!}{
\begin{tabular}{ccccc}
\hline
\textbf{Element} & $\bs{k}$ [W/(m K)] & \textbf{Thickness} [m] & \textbf{Density} [kg/m$^3$] & $\bs{c_p}$ [J/(kg K)]\\
\hline
\hline
\multicolumn{5}{c}{\textbf{Exterior wall (inside to outside)}}\\
Plasterboard     & $0.16$ & $0.012$  & $950$ & $840$\\
Fiberglass quilt & $0.04$ & $0.066$  & $12$  & $840$\\
Wood slicing     & $0.14$ & $0.009$  & $530$ & $900$\\
\hline
\multicolumn{5}{c}{\textbf{Floor (bottom to up)}}\\
Timbering floor  & $0.14$ & $0.025$  & $650$ & $1200$\\
Insulation       & $0.04$ & $1.003$  & $0$   & $0$   \\
\hline
\multicolumn{5}{c}{\textbf{Roof (inside to outside)}}\\
Plasterboard     & $0.16$ & $0.1$    & $950$ & $840$\\
Fiberglass quilt & $0.04$ & $0.1118$ & $12$  & $840$\\
Roofdeck         & $0.14$ & $0.019$  & $530$ & $900$\\
\hline
\end{tabular}
}
\end{table}

\begin{table}[htb]
\centering
\caption{Walls composition for the heavyweight building case. }
\label{tab:material_heavyweight}
\resizebox{\textwidth}{!}{
\begin{tabular}{ccccc}
\hline
\textbf{Element} & $\bs{k}$ [W/(m K)] & \textbf{Thickness} [m] & \textbf{Density} [kg/m$^3$] & $\bs{c_p}$ [J/(kg K)]\\
\hline
\hline
\multicolumn{5}{c}{\textbf{Exterior wall (inside to outside)}}\\
Concrete block   & $0.51$ & $0.1$    & $1400$& $1000$\\
Foam insulation  & $0.04$ & $0.0615$ & $10$  & $1400$\\
Wood slicing     & $0.14$ & $0.009$  & $530$ & $900$\\
\hline
\multicolumn{5}{c}{\textbf{Floor (bottom to up)}}\\
Concrete slab    & $1.13$ & $0.08$   & $1400$& $1000$\\
Insulation       & $0.04$ & $1.007$  & $0$   & $0$   \\
\hline
\multicolumn{5}{c}{\textbf{Roof (inside to outside)}}\\
Plasterboard     & $0.16$ & $0.1$    & $950$ & $840$\\
Fiberglass quilt & $0.04$ & $0.1118$ & $12$  & $840$\\
Roofdeck         & $0.14$ & $0.019$  & $530$ & $900$\\
\hline
\end{tabular}
}
\end{table}

The standard also provides the values for the internal and external solar absorption and infrared emission coefficients
 $\alpha_i^S = \alpha_i^L = 0.6$ and $\varepsilon_i = 0.9$, $i=1,\dots,m$, and for the interior and exterior combined radiative and convective heat transfer coefficients, from which the radiative and convective coefficients can be recovered. Finally, the standard contains also the windows properties, the values of incidence angle-dependent optical properties, and the interior solar distribution. The reader is referred to the ANSI-ASHRAE 140 standard  for a complete list of building properties.

We focus on two procedures for validation described in the standard. The first one is denoted as Free Float (FF) in that the heating and cooling equipment is switched off and the zone temperature evolves freely subject to internal/external disturbances. The purpose of this test is to validate the physical model without the effect of any control action, and such validation is performed comparing some statistics (maximum, minimum, and annual average) of the zone temperature $T_z$ over a year against other simulation tools. The second procedure prescribes to simulate the building together with the heating/cooling system by applying a simple control strategy: the controller has to maintain the air temperature inside the building between $20^\circ$C and $27^\circ$C. Specifically, the control strategy is:
\begin{itemize}
\item Heat = ON if temperature $<20^\circ$C; otherwise, Heat = OFF.
\item Cool = ON if temperature $>27^\circ$C; otherwise, Cool = OFF.
\end{itemize}
The ANSI-ASHRAE 140 standard specifies that the air conditioning system produces only pure heating load and sensible cooling load outputs. That is, all equipment is $100$\% efficient with no duct losses and no capacity limitations. In this controlled case, the validation is performed comparing the hourly-integrated peak of the cooling and heating power provided to the building.
The thermostat was implemented as two saturated PI controllers with antiwindup, where the control variable is the amount of cooling and heating power to be injected in the system, equivalently to the implementation adopted in~\cite{wetter2014jbps}. 

In the following we will denote as 600FF and 900FF the case when the free float validation procedure is applied to the lightweight and heavyweight buildings, and as 600 and 900 the case when the control is applied.

\subsection{Validation results}

In this section we present the numerical results obtained in the 600FF and 900FF and 600 and 900 test cases.
For running the validation process, it is necessary to rewrite the model with the heat flow rate $Q$ as control input, and the temperature of the zone $T_z$ as the output of the system. To this aim, we consider a simulation model composed of a state vector including the temperature of the different slices of the walls as described in~\eqref{eq:building_dynamics}, and the temperature of the zone $T_z$. The evolution of $T_z$ is governed by the continuous-time version of~\eqref{eq:zone_energy_equation}, made explicit with respect to $\dot T_z$:
\begin{align*}
\dot{T}_z = -C_z^{-1}Q_z = -C_z^{-1}(Q - Q_w - Q_p - Q_{\text{int}}),
\end{align*}
with $Q_w$, $Q_p$, and $Q_{\text{int}}$ being the heat flow rates towards the zone of the walls, the occupancy, and of other internal equipment producing heat. Considering the expressions~\eqref{eq:building_output},~\eqref{eq:people_heat_rate_lin}, and~\eqref{eq:windows_heat_rate}, one can write the expression of $Q_z$ as a function of the states $T$ and $T_z$, of the input $Q$ and of the disturbances. This continuous time model is implemented in Modelica\footnote{\url{https://modelica.org/}}, in order to carry out the validation process.

Table~\ref{tab:validationResultsFF} reports the obtained results in terms of maximum, minimum, and mean annual temperature. 
Our model (last column of the table) is compared  with the other ones provided in the standard under the free float validation procedure.
\begin{table}[h]
\centering
\caption{Comparative analysis results for the free float experiments.}
\label{tab:validationResultsFF}
\resizebox{\textwidth}{!}{
\begin{tabular}{cccccccccc}
\hline
\textbf{Case} & \textbf{ESP} & \textbf{BLAST} & \textbf{DOE2} & \textbf{SRES} & \textbf{SERIRES} & \textbf{S3PAS} & \textbf{TRNSYS} & \textbf{TASE} & \textbf{Modelica}\\
\hline
\hline
\multicolumn{10}{c}{\textbf{Maximum temperature [$^\circ$C]}}\\
\hline
600FF & $64.9 $ & $65.1 $ & $69.5 $ & $68.8 $ & --     & $64.9 $ & $65.3 $ & $65.3 $ & $65.96 $\\
900FF & $41.8 $ & $43.4 $ & $42.7 $ & $44.8 $ & --     & $43.0 $ & $42.5 $ & $43.2 $ & $47.09 $\\
\hline
\multicolumn{10}{c}{\textbf{Minimum temperature [$^\circ$C]}}\\
\hline
600FF & $-15.8$ & $-17.1$ & $-18.8$ & $-18.0$ & --     & $-17.8$ & $-17.8$ & $-18.5$ & $-21.48$\\
900FF & $-1.6 $ & $-3.2 $ & $-4.3 $ & $-4.5 $ & --     & $-4.0 $ & $-6.4 $ & $-5.6 $ & $-3.17 $\\
\hline
\multicolumn{10}{c}{\textbf{Mean annual temperature [$^\circ$C]}}\\
\hline
600FF & $25.1 $ & $25.4 $ & $24.6 $ & $25.5 $ & $25.9$ & $25.2 $ & $24.5 $ & $24.2 $ & $25.62 $\\
900FF & $25.5 $ & $25.9 $ & $24.7 $ & $25.5 $ & $25.7$ & $25.2 $ & $24.5 $ & $24.5 $ & $21.81 $\\
\hline
\end{tabular}
}
\end{table}

In the 600FF test case, the results obtained with the model considered herein are comparable with the ones obtained with the other building simulation models. As for 900FF, only the minimum temperature is comparable with the other results, while the maximum temperature is slightly higher than the values obtained with the other models, whereas the mean annual temperature is lower. Overall the obtained statistics produce reasonable results in the free float case, even though the model adopted for the presented framework is much simpler than the other simulation models.

Table~\ref{tab:validationResultsControlled} summarizes the validation results when the presented control strategy is in place.
\begin{table}[h]
\centering
\caption{Hourly integrated peak of the heating and cooling power provided to the building for test cases 600 and 900.}
\label{tab:validationResultsControlled}
\resizebox{\textwidth}{!}{
\begin{tabular}{cccccccccc}
\hline
\textbf{Case} & \textbf{ESP} & \textbf{BLAST} & \textbf{DOE2} & \textbf{SRES} & \textbf{SERIRES} & \textbf{S3PAS} & \textbf{TRNSYS} & \textbf{TASE} & \textbf{Modelica}\\
\hline
\hline
\multicolumn{10}{c}{\textbf{Heating [kW]}}\\
\hline
600 & $3.437$ & $3.940$ & $4.045$ & $4.258$ & --     & $4.037$ & $3.931$ & $4.354$ & $4.521$\\
900 & $2.850$ & $3.453$ & $3.557$ & $3.760$ & --     & $3.608$ & $3.517$ & $3.797$ & $4.077$\\
\hline
\multicolumn{10}{c}{\textbf{Cooling [kW]}}\\
\hline
600 & $6.194$ & $5.965$ & $6.656$ & $6.627$ & --     & $6.286$ & $6.488$ & $6.812$ & $6.983$\\
900 & $2.888$ & $3.155$ & $3.458$ & $3.871$ & --     & $3.334$ & $3.567$ & $3.457$ & $3.922$\\
\hline
\end{tabular}
}
\end{table}

The hourly peak of cooling and heating power are comparable with those of the other tools in both the test cases.\\
Overall, the obtained results in this validation phase show that state-of-the-art simulation tools provide similar results to those obtained with our model, which has the key advantage of being simpler and hence more suitable for design purposes. 

 \section{Discretization of the walls temperature dynamics}
\label{sec:discretization_walls}

To solve the discrete-time optimal energy management problem, we need to consider a discretized version of \eqref{eq:building_state_space}. Given the linearity of \eqref{eq:building_state_space}, it holds that
\begin{align}
\bs{T}((k+1)\dt) &= e^{\bs{A}\dt}\bs{T}(k\dt) +  \int_{k\dt}^{(k+1)\dt} e^{\bs{A}((k+1)\dt-\tau)}(\bs{B}\bs{T_z}(\tau) + \bs{W}\bs{d}(\tau)) d\tau.
\label{eq:discr_state_movement}
\end{align}
If we assume that $\bs{T_z}$ (i.e., our control variables) and $\bs{d}$ are linearly varying within each time slot, then the integral in \eqref{eq:discr_state_movement} can be computed analytically. Formally, given
\begin{equation}
	\bs{T_z}(\tau) = \frac{\bs{T}_{\bs{z},k+1}-\bs{T}_{\bs{z},k}}{\dt}(\tau - k\dt) + \bs{T}_{\bs{z},k},
	\label{eq:discr_temp}
\end{equation}
where $\bs{T}_{\bs{z},k} = \bs{T_z}(k\dt)$, and
\begin{equation*}
	\bs{d}(\tau) = \frac{\bs{d}_{k+1}-\bs{d}_k}{\dt}(\tau - k\dt) + \bs{d}_k,
\end{equation*}
where $\bs{d}_k = \bs{d}(k\dt)$, $\forall \tau: \; k\dt \leq \tau < (k+1)\dt$ and $k=1,\dotsc,M$. If we set $\bs{T}_k = \bs{T}(k\dt)$ and $\bs{Q}_k = \bs{Q}(k\dt)$, $k=1,\dotsc,M$, then the dicretized system can be expressed as follows
\begin{equation*}
\begin{cases}
\begin{aligned}
	\bs{T}_{k+1} &= \bs{\Gamma_x}\bs{T}_k + \bs{\Gamma_{u,1}}\bs{T}_{\bs{z},k+1}
			+ (\bs{\Gamma_{u,0}}-\bs{\Gamma_{u,1}})\bs{T}_{\bs{z},k} + \\
			& \qquad \qquad \! + \bs{\Gamma_{\omega,1}}\bs{d}_{k+1} + (\bs{\Gamma_{\omega,0}}-\bs{\Gamma_{\omega,1}})\bs{d}_k \\
	\bs{Q}_k &= \; \bs{C}\bs{T}_k \;\! + \bs{D}\bs{T}_{\bs{z},k} \\
\end{aligned}
\end{cases}
\end{equation*}
where
\begin{align*}
&\bs{\Gamma_x} = e^{\bs{A}\dt}\\
&\bs{\Gamma_{u,1}} = \cfrac{1}{\dt} \left(\int_0^{\dt} e^{\bs{A}s}(\dt-s)ds\right) \bs{B} \\
&\bs{\Gamma_{u,0}} = \left(\int_0^{\dt} e^{\bs{A}s} ds\right) \bs{B}\\
&\bs{\Gamma_{\omega,1}} = \cfrac{1}{\dt} \left(\int_0^{\dt} e^{\bs{A}s}(\dt-s)ds\right) \bs{W}\\
&\bs{\Gamma_{\omega,0}} = \left(\int_0^{\dt} e^{\bs{A}s} ds\right) \bs{W}.
\end{align*}

Applying the transformation $\bs{\xi}_k = \bs{T}_k-\bs{\Gamma_{u,1}}\bs{T}_{\bs{z},k}-\bs{\Gamma_{\omega,1}}\bs{d}_k$ we
obtain
\begin{equation}
\begin{cases}
\begin{aligned}
	\bs{\xi}_{k+1} &= \bs{\Gamma_x}\bs{\xi}_k
		+ ((\bs{\Gamma_x}-\bs{I})\bs{\Gamma_{u,1}} + \bs{\Gamma_{u,0}})\bs{T}_{\bs{z},k} +\\
		& \qquad \quad \;\;\;\! + ((\bs{\Gamma_x}-\bs{I})\bs{\Gamma_{\omega,1}} +\bs{\Gamma_{\omega,0}})\bs{d}_k \\
	\bs{Q}_k &= \; \bs{C}\bs{\xi}_k \;\! + (\bs{C}\bs{\Gamma_{u,1}} + \bs{D})\bs{T}_{\bs{z},k}
		+ \bs{C}\bs{\Gamma_{\omega,1}} \bs{d}_k \\
\end{aligned}
\end{cases}
\label{eq:building_discr_trasl}
\end{equation}
Dropping the bold notation for vectors and matrices, \eqref{eq:building_discr_trasl} can be rewritten as the following discrete-time system
\begin{equation}
		\left\{
		\begin{aligned}
			&x(k+1) = \tilde{A}x(k) + \tilde{B}u(k) + \tilde{W}\omega(k) \\
			&y(k) = \tilde{C}x(k) + \tilde{D}u(k) + \tilde{V}\omega(k) \\
		\end{aligned}
		\right.
	\label{eq:discrete_general}
\end{equation}
where $x(k) = \bs{\xi}_k$, $u(k) = \bs{T}_{\bs{z},k}$, $\omega(k) = \bs{d}_k$, $y(k) = \bs{Q}_k$, and the matrices are:
\begin{equation*}
\begin{array}{lll}
\tilde{A} = \bs{\Gamma_x}, \quad & \tilde{B} = (\bs{\Gamma_x}-\bs{I})\bs{\Gamma_{u,1}}+\bs{\Gamma_{u,0}}, \quad & \tilde{W} = (\bs{\Gamma_x}-\bs{I})\bs{\Gamma_{\omega,1}}+\bs{\Gamma_{\omega,0}}\\
\tilde{C} = \bs{C}, \quad &\tilde{D} = \bs{C}\bs{\Gamma_{u,1}} + \bs{D}, \quad &\tilde{V} = \bs{C}\bs{\Gamma_{\omega,1}}.
\end{array}
\end{equation*}

From \eqref{eq:discrete_general} one can derive the expression of $x(k)$ and $y(k)$ as a function of the initial state and the inputs up to  $k$:
\begin{align*}
 	&x(k) = \tilde{A}^k x(0) + \sum_{h=0}^{k-1} \tilde{A}^{k-1-h} \left( \tilde{B}u(h)+\tilde{W}\omega(h) \right) \\
 	&y(k) = \tilde{C}\tilde{A}^k x(0) + \sum_{h=0}^{k-1} \tilde C\tilde{A}^{k-1-h} \left( \tilde{B}u(h)+\tilde{W}\omega(h) \right) + \tilde{D}u(k) + \tilde{V}\omega(k).
\end{align*}

By recalling that $x(0)=\bs{\xi}_0 = \bs{T}_0-\bs{\Gamma_{u,1}}\bs{T}_{\bs{z},0}-\bs{\Gamma_{\omega,1}}\bs{d}_0$, we obtain the following expression of $y(k)$ as a function of the original state variables and the inputs up to time $k$:
\begin{align*}
	y(k) =
&\, \tilde{C}\tilde{A}^{k}\tilde{A} \bs{T}_0+\tilde{C}\tilde{A}^{k-1}\left( (\bs{\Gamma_{u,0}}-\bs{\Gamma_{u,1}})u(0) + (\bs{\Gamma_{\omega,0}}-\bs{\Gamma_{\omega,1}})\omega(0) \right) \\
		&+ \sum_{h=1}^{k-1} \tilde C\tilde{A}^{k-1-h} \left( \tilde{B}u(h)+\tilde{W}\omega(h) \right) + \tilde{D}u(k) + \tilde{V}\omega(k).
\end{align*}

\bibliographystyle{elsarticle-num}
\bibliography{microgrid}

\end{document}